\documentstyle[12pt,preprint,tighten,floats,aps,epsf,psfig]{revtex}
\begin{document}
\setlength{\unitlength}{1mm}
\textwidth 16.3 true cm
\textheight 23.0 true cm
\topmargin -0.1 true in
\oddsidemargin 0.00 true in
\setlength{\headheight}{0in}
\setlength{\headsep}{0in}
\setlength{\topskip}{1ex}
\setlength{\textheight}{8.5in}
\setlength{\textwidth}{7.25in}
\setlength{\topmargin}{0.5cm}
\setlength{\oddsidemargin}{0.25in}
\setlength{\evensidemargin}{\oddsidemargin}
\setlength{\baselineskip}{0.24in}
\newcommand{\newc}{\newcommand}
\def\be{\begin{equation}}
\def\ee{\end{equation}}
\def\bea{\begin{eqnarray}}
\def\eea{\end{eqnarray}}
\def\simlt{\stackrel{<}{{}_\sim}}
\def\simgt{\stackrel{>}{{}_\sim}}
\def\beq{\begin{equation}}
\def\eeq{\end{equation}}
\def\bea{\begin{eqnarray}}
\def\eea{\end{eqnarray}}
\def\ve{\vert}
\def\vel{\left|}
\def\ver{\right|}
\def\nnb{\nonumber}
\def\ga{\left(}
\def\dr{\right)}
\def\aga{\left\{}
\def\adr{\right\}}
\def\rar{\rightarrow}
\def\nnb{\nonumber}
\def\la{\langle}
\def\ra{\rangle}
\def\ba{\begin{array}}
\def\ea{\end{array}}
\def\tep{$B \rar K \ell^+ \ell^-$}
\def\tepm{$B \rar K \mu^+ \mu^-$}
\def\tept{$B \rar K \tau^+ \tau^-$}
\def\ds{\displaystyle}
\def\beq{\begin{equation}}
\def\eeq{\end{equation}}
\def\bea{\begin{eqnarray}}
\def\eea{\end{eqnarray}}
\def\ve{\vert}
\def\vel{\left|}
\def\ver{\right|}
\def\nnb{\nonumber}
\def\ga{\left(}
\def\dr{\right)}  
\def\aga{\left\{} 
\def\adr{\right\}}
\def\rar{\rightarrow}
\def\nnb{\nonumber}
\def\la{\langle}
\def\ra{\rangle}
\def\lla{\left<}
\def\rra{\right>}
\def\ba{\begin{array}}
\def\ea{\end{array}}
\def\tep{$B \rar K \ell^+ \ell^-$}
\def\tepm{$B \rar K \mu^+ \mu^-$}
\def\tept{$B \rar K \tau^+ \tau^-$}
\def\ds{\displaystyle}
\def\lesssim{\mathrel{\mathpalette\vereq<}}
\def\vereq#1#2{\lower3pt\vbox{\baselineskip1.5pt \lineskip1.5pt
\ialign{$\m@th#1\hfill##\hfil$\crcr#2\crcr\sim\crcr}}}
\def\gtrsim{\mathrel{\mathpalette\vereq>}}
\def\alt{\lesssim}
\def\agt{\gtrsim}
\def\bos{\lower 0.5cm\hbox{{\vrule width 0pt height 1.2cm}}}
\def\boss{\lower 0.35cm\hbox{{\vrule width 0pt height 1.cm}}}
\def\aaa{\lower 0.cm\hbox{{\vrule width 0pt height .7cm}}}
\def\dol{\lower 0.4cm\hbox{{\vrule width 0pt height .5cm}}}
\title{Probing the sources of CP violation via $B\rightarrow K^{\star} \ell^{+}\ell^{-}$ decay}
\author{T. M. Aliev$^{a}$, D. A. Demir$^{b}$, M. Savc{\i}$^{a}$}
\address{$^a$ Middle East Technical University, Department of Physics, 06531 Ankara, Turkey\\
$^b$ The Abdus Salam International Centre for Theoretical Physics, I-34100, Trieste, Italy}
\maketitle
\begin{abstract}
The $B \rar K^\ast \ell^-\ell^+$ ($\ell = \mu,~\tau$) is analyzed in 
a minimally extended Standard Model in which the Wilson coefficients
have new CP--odd phases. The sensitivity of the CP asymmetry and 
lepton polarization asymmetries to the new phases is discussed. 
It is found that the CP asymmetry is sensitive to the new phase
in the Wilson coefficient $C_7$ whereas the normal lepton
polarization asymmetry is sensitive to the phase in the Wilson 
coefficient $C_{10}$. Additionally, the correlation between the
CP and normal lepton polarization asymmetries is studied. A simultaneous 
measurement of these two asymmetries can be useful in search for 
the existence of the new sources of CP violation beyond the Standard Model.
\end{abstract}
\newpage
\section{Introduction}
Violation of the CP symmetry has now become a well--established fact 
in Kaon system \cite{kaon}. Once the era of
$B$--factories start with the operation of KEK-B, B-TeV, LHC-B and SLAC's
Asymmetric $B$ factory, it will be possible to test the standard model 
(SM) at one--loop accuracy. In general, possible incompatibility of the
experimental data with the SM predictions will mark the existence of
`new physics' contributions.  Among all, an experimental determination
of the CP--violating quantities and their comparison with the SM predictions 
will be particularly useful in search for the new physics effects.

From the experimental perspective the exclusive decay modes 
(such as $B\rightarrow K^\ast \gamma$ \cite{CLEO}, $B\rightarrow K^\ast \ell^+
\ell^-$, $B\rightarrow K \ell^+ \ell^-$) are easy to measure. From 
the theoretical view point, however, the corresponding inclusive
modes ($b\rightarrow s \gamma$ and $b\rightarrow s \ell^+ \ell^-$) 
can be cleanly estimated as the only machinery needed are the Wilson 
coefficients describing the short--distance physics. 
A proper description of the exclusive decay modes, on the other hand, 
depends on both Wilson coefficients (short--distance physics) 
and the hadronic form factors (long--distance physics). This causes 
a relative increase of the uncertainties due to hadronization 
effects.

For the purpose of studying the sources of CP violation, it is 
convenient to concentrate on those observables which are sensitive to
the possible CP phases. Among these, for example, CP asymmetries
and lepton polarization asymmetries are such ones \cite{R2}--\cite{R4}. 
Recently, a detailed study of the lepton polarization asymmetries in 
$B\rightarrow X_{s} \ell^{+} \ell^{-}$ decay has been performed in
a rather general model by including nine additional Wilson coefficients 
not found in the SM \cite{R5}. Keeping this kind of short--distance 
generality it is convenient to discuss the exclusive decay 
modes such as $B\rightarrow K^{\ast} \ell^{+} \ell^{-}$ \cite{R6,R7}.
Such an analysis will be useful for a first--hand comparison with 
the experiment as the inclusive modes are generally hard to measure. 

In Sec. II we start with a general non--standard description of the 
short--distance physics as in \cite{R5}. Then we parametrize the
long--distance quantities by appropriate form factors and obtain
the hadronic transition amplitude. In Sec. III we derive general
analytic expressions for asymmetries will be give. In doing this
all sources of CP violation will be ascribed to short--distance 
physics. In Sec. IV the asymmetries and their relation 
to the Wilson coefficients will be analyzed numerically. 
In Sec. V results are discussed and the conclusion is stated.

\section{The Decay Amplitude}
The exclusive $B$ decays $B\rightarrow K^{\ast} \ell^{+}\ell^{-}$ are 
conveniently described by the partonic decay $b\rightarrow s \ell^{+}\ell^{-}$
at distances ${\cal{O}}(M_{W}^{-1})$. The effective Hamiltonian describing this 
rare $b$ decay at the scale $\mu \sim M_{W}$ should, however, be evolved down to 
mesonic mass scale $\mu\sim m_{b}$ using the QCD evolution equations. Then 
the decay amplitude describing $b\rightarrow s \ell^{+}\ell^{-}$ takes the form \cite{R5,R6}
\bea
\label{partoamp}
\mbox{M}(b\rightarrow s \ell^{+}\ell^{-}) &=& \frac{G\alpha}{\sqrt{2} \pi}
 V_{tb}V_{ts}^\ast
\Bigg\{ C_{SL} \, \bar s i \sigma_{\mu\nu} \frac{q^\nu}{q^2} (m_s L) b   
\, \bar \ell \gamma_\mu \ell + C_{BR}\, \bar s i \sigma_{\mu\nu}
\frac{q^\nu}{q^2} (m_b R) b \, \bar \ell \gamma_\mu \ell \nnb \\
&+&C_{LL}\, \bar s_L \gamma_\mu b_L \,\bar \ell_L \gamma^\mu \ell_L +   
C_{LR} \,\bar s_L \gamma_\mu b_L \, \bar \ell_R \gamma^\mu \ell_R +
C_{RL} \,\bar s_R \gamma_\mu b_R \,\bar \ell_L \gamma^\mu \ell_L \nnb \\
&+&C_{RR} \,\bar s_R \gamma_\mu b_R \, \bar \ell_R \gamma^\mu \ell_R +
C_{LRLR} \, \bar s_L b_R \,\bar \ell_L \ell_R +
C_{RLLR} \,\bar s_R b_L \,\bar \ell_L \ell_R \nnb \\
&+&C_{LRRL} \,\bar s_L b_R \,\bar \ell_R \ell_L +
C_{RLRL} \,\bar s_R b_L \,\bar \ell_R \ell_L+
C_T\, \bar s \sigma_{\mu\nu} b \,\bar \ell \sigma^{\mu\nu}\ell \nnb \\
&+&i C_{TE}\,\epsilon^{\mu\nu\alpha\beta} \bar s \sigma_{\mu\nu} b \,
\bar \ell \sigma_{\alpha\beta} \ell  \Bigg\}~,
\eea
where each of the Wilson coefficients $C_{SL},\cdots, C_{TE}$ is evaluated at the $B$-meson mass scale,
$\mu\sim m_{b}$. In this expression, $L(R)=(1-(+)\gamma_{5})/2$ are the left (right) projection operators,
$V_{ij}$ are the elements of the CKM matrix, and $q=p_{B}-p_{K^\ast}=p_{+}+p_{-}$ is the 
momentum transfer to the dilepton channel. This decay amplitude has a rather general form as 
it includes nine additional operators not found in the minimal standard model. The only simplifying 
assumption about this decay amplitude will be twofold: (1) Neglect of the strange quark mass 
everywhere in the analysis, (2) Neglect of the tensor operators having the coefficients $C_{T}$ and $C_{TE}$.
The former is justified by the smallness of the ratio $m_{s}/m_{b}$ and the latter is justified 
by the previous analyzes which show that their contributions are much smaller than other 
operators (for details, see \cite{R9}). 

The quark level decay amplitude (\ref{partoamp}) controls the
semileptonic decays $B\rightarrow \left( K, K^{*}\right) \ell^{+}\ell^{-}$.
The amplitudes for these exclusive decays can be obtained after evaluating 
the matrix elements of the quark operators in (\ref{partoamp}) between the
$\left.|B(p_{B})\rra$ and $\lla K^\ast(p_{K^\ast})|\right.$ states. In particular, explicit expressions for
$\lla K^\ast \vel \bar s \gamma_\mu (1\pm \gamma_5) b \ver B \rra$, $\lla K^\ast \vel \bar s i \sigma_{\mu\nu} q^\nu (1+
\gamma_5) b \ver B \rra$ and $\lla K^\ast \vel \bar s (1\pm \gamma_5) b \ver B \rra$ are needed. Computation of such
hadronic matrix elements is bound to parametrizations of the form factors depending only on the momentum transfer
square, or equivalently, the dilepton invariant mass $m_{\ell\ell}^{2}=(p_{B}-p_{K^{\ast}})^{2}=(p_{+}+p_{-})^{2}\equiv q^{2}$.
Introducing appropriate from factors one obtains
\bea
\lefteqn{
\label{ilk}
\lla K^\ast(p_{K^\ast},\varepsilon) \vel \bar s \gamma_\mu (1 \pm \gamma_5) b \ver
B(p_B) \rra =} \nnb \\
&&- \epsilon_{\mu\nu\rho\sigma} \varepsilon^{\ast\nu} p_{K^\ast}^\rho q^\sigma
\frac{2 V(q^2)}{m_B+m_{K^\ast}} \pm i \varepsilon_\mu^\ast (m_B+m_{K^\ast})
A_1(q^2) \mp i (p_B + p_{K^\ast})_\mu (\varepsilon^\ast q)
\frac{A_2(q^2)}{m_B+m_{K^\ast}} \nnb \\
&&\mp i q_\mu \frac{2 m_{K^\ast}}{q^2} (\varepsilon^\ast q)
\left[A_3(q^2)-A_0(q^2)\right]~, \nnb \\ \\
\lefteqn{
\label{iki}
\lla K^\ast(p_{K^\ast},\varepsilon) \vel \bar s i \sigma_{\mu\nu} q^\nu
(1 + \gamma_5) b \ver B(p_B) \rra =} \nnb \\
&&4 \epsilon_{\mu\nu\rho\sigma} \varepsilon^{\ast\nu} p_{K^\ast}^\rho q^\sigma
T_1(q^2) + 2 i \left[ \varepsilon_\mu^\ast (m_B^2-m_{K^\ast}^2) -
(p_B + p_{K^\ast})_\mu (\varepsilon^\ast q) \right] T_2(q^2) \nnb \\
&&+ 2 i (\varepsilon^\ast q) \left[ q_\mu -
(p_B + p_{K^\ast})_\mu \frac{q^2}{m_B^2-m_{K^\ast}^2} \right] T_3(q^2)~,
\eea
where the explicit expressions for $V(q^2)$, $A_{0,1,2,3}(q^{2})$ and $T_{1,2,3}(q^{2})$ 
will be given below.

To ensure the finiteness of (\ref{ilk}) as $q^2\rightarrow 0$, it is usually assumed that
$A_3(q^2=0) = A_0(q^2=0)$. Besides, to calculate the matrix elements of
the scalar operators, $\lla K^\ast \vel \bar s (1 \pm \gamma_5) b \ver B \rra$,
it is necessary to contract (\ref{ilk}) with $q_\mu$ and use the equation of motion, giving 
\bea
\lefteqn{
\label{uc}
\lla K^\ast(p_{K^\ast},\varepsilon) \vel \bar s (1 \pm \gamma_5) b \ver
B(p_B) \rra =
\frac{1}{m_b} \Big\{ \mp i (\varepsilon^\ast q) (m_B+m_{K^\ast})
A_1(q^2)}\nnb \\
&&~~~~~~~~~~~~~\pm i (m_B-m_{K^\ast}) (\varepsilon^\ast q) A_2(q^2)
\pm 2 i m_{K^\ast} (\varepsilon^\ast q) \left[A_3(q^2)-A_0(q^2)\right]~\Big\}.
~~~~~~~~~~~
\eea
Additionally, again using the equation of motion, the form factor $A_3$ can be expressed as a
linear combination of the form factors $A_1$ and $A_2$ (see \cite{R10})
\bea
A_3(q^2) = \frac{m_B+m_{K^\ast}}{2 m_{K^\ast}} A_1(q^2) -
\frac{m_B-m_{K^\ast}}{2 m_{K^\ast}} A_2(q^2)~.
\eea
Having this relation at hand, one finally obtains
\bea
\label{dort}
\lla K^\ast(p_{K^\ast},\varepsilon) \vel \bar s (1 \pm \gamma_5) b \ver B(p_B) \rra
= \frac{1}{m_b} \Big\{\mp 2 i m_{K^\ast} (\varepsilon^\ast q) A_0(q^2)
\Big\}~.
\eea
This completes the evaluation of the necessary transition matrix elements. 

As mentioned before the form factors entering (\ref{ilk})-(\ref{dort}) represent the
hadronization process which lacks a Lagrangian description. They are thus generally 
computed in framework of certain nonperturbative approaches  such as chiral theory 
\cite{R11}, three point QCD sum rules method \cite{R10}, relativistic quark model 
by the light--front formalism \cite{R12}, effective heavy quark theory 
\cite{R13} and light cone QCD sum rules \cite{R14,R15,R16}. In what follows 
we will use the results of the work \cite{R15} in which the form factors are described by a
three--parameter fit where the radiative corrections up to leading twist contribution and 
SU(3)--breaking effects are taken into account. Letting 
$F(q^{2})\in\{V(q^2), A_{0}(q^{2}), A_{1}(q^{2}), A_{2}(q^{2}), A_{3}(q^{2}), T_{1}(q^{2}), T_{2}(q^{2}),
T_{3}(q^{2})\}$, the $q^{2}$--dependence of any of these form factors could be
parametrized as 
\bea
\label{formfac}
F(s) = \frac{F(0)}{1-a_F\,s + b_F\, s^{2}}~, \nnb
\eea
where the parameters $F(0)$, $a_F$ and $b_F$ are listed in Table 1 for each form factor. Here $s=q^2/m_{B}^{2}$
is the dilepton invariant mass in units of $B$--meson mass (See, \cite{R15,R16}). 

\begin{table}[h]
\renewcommand{\arraystretch}{1.5}
\addtolength{\arraycolsep}{3pt}
$$
\begin{array}{|l|ccc|}
\hline
& F(0) & a_F & b_F \\ \hline
A_0^{B \rar K^*} &\phantom{-}0.47 & 1.64 & 0.94 \\
A_1^{B \rar K^*} &
\phantom{-}0.34 \pm 0.05 & 0.60 & -0.023 \\
A_2^{B \rar K^*} &
\phantom{-}0.28 \pm 0.04 & 1.18 & \phantom{-}0.281\\
V^{B \rar K^*} &
 \phantom{-}0.46 \pm 0.07 & 1.55 & \phantom{-}0.575\\
T_1^{B \rar K^*} &
  \phantom{-}0.19 \pm 0.03 & 1.59 & \phantom{-}0.615\\
T_2^{B \rar K^*} &
 \phantom{-}0.19 \pm 0.03 & 0.49 & -0.241\\
T_3^{B \rar K^*} &
 \phantom{-}0.13 \pm 0.02 & 1.20 & \phantom{-}0.098\\ \hline
\end{array}
$$
\caption{The form factors for $B\rightarrow K^\ast \ell^{+}\ell^{-}$
in a three--parameter fit.}
\renewcommand{\arraystretch}{1}
\addtolength{\arraycolsep}{-3pt}
\end{table}

Making use of the hadronic matrix elements (\ref{ilk})-(\ref{dort}) of the basic quark current structures in 
(\ref{partoamp}), it is straightforward to determine the decay amplitude for $B \rar K^\ast \ell^+ \ell^-$ decay:
\bea
\lefteqn{
\label{hadroamp}
\mbox{M}(B\rightarrow K^\ast \ell^{+}\ell^{-}) =
\frac{G \alpha}{4 \sqrt{2} \pi} V_{tb} V_{ts}^\ast }\nnb \\
&&\times \Bigg\{
\bar \ell \gamma_\mu(1-\gamma_5) \ell \, \Big[
-2 {\cal{V}}_{L_1}\epsilon_{\mu\nu\rho\sigma} \varepsilon^{\ast\nu} p_{K^\ast}^\rho q^\sigma
 -i{\cal{V}}_{L_2}\varepsilon_\mu^\ast
+ i {\cal{V}}_{L_3}(\varepsilon^\ast q) (p_B+p_{K^\ast})_\mu
+ i {\cal{V}}_{L_4}(\varepsilon^\ast q) q_\mu  \Big] \nnb \\
&&+ \bar \ell \gamma_\mu(1+\gamma_5) \ell \, \Big[
-2 {\cal{V}}_{R_1}\epsilon_{\mu\nu\rho\sigma} \varepsilon^{\ast\nu} p_{K^\ast}^\rho q^\sigma
 -i{\cal{V}}_{R_2}\varepsilon_\mu^\ast    
+ i {\cal{V}}_{R_3}(\varepsilon^\ast q) (p_B+p_{K^\ast})_\mu
+ i {\cal{V}}_{R_4}(\varepsilon^\ast q) q_\mu  \Big] \nnb \\
&&+\bar \ell (1-\gamma_5) \ell \Big[ i {\cal{S}}_{L}(\varepsilon^\ast q)\Big] 
+ \bar \ell (1+\gamma_5) \ell \Big[ i {\cal{S}}_{R}(\varepsilon^\ast q)\Big]\Bigg\}~,
\eea
where ${\cal{V}}_{L_i}$ and ${\cal{V}}_{R_i}$ are the coefficients of left-- and right--handed
leptonic currents with vector structure, respectively. Clearly, ${\cal{S}}_{L,R}$ are the weights of
scalar leptonic currents with respective chirality. These new coefficients are functions of the Wilson
coefficients in the partonic decay amplitude (\ref{partoamp}) and the form factors introduced in defining
the hadronic matrix elements above. Their explicit expressions are given by
\bea
{\cal{V}}_{L_1} &=& (C_{LL} + C_{RL}) \frac{V(q^2)}{m_B+m_{K^\ast}} -
2 C_{BR} \frac{m_b}{q^2} \, T_1 , \nnb \\
{\cal{V}}_{L_2} &=& (C_{LL} - C_{RL}) (m_B+m_{K^\ast}) A_1 - 2 C_{BR}
\frac{m_b}{q^2} (m_B^2-m_{K^\ast}^2) \, T_2~, \nnb \\
{\cal{V}}_{L_3} &=& \frac{C_{LL} - C_{RL}}{m_B+m_{K^\ast}} A_2 - 2 C_{BR}
\frac{m_b}{q^2}  \left[ T_2 + \frac{q^2}{m_B^2-m_{K^\ast}^2} T_3 \right]~,
\nnb \\
{\cal{V}}_{L_4} &=& (C_{LL} - C_{RL}) \frac{2 m_{K^\ast}}{q^2} (A_3-A_0)-
2 C_{BR} \frac{m_b}{q^2} T_3~, \nnb \\
{\cal{V}}_{R_1} &=& {\cal{V}}_{L_1} ( C_{LL} \rar C_{LR}~,~~C_{RL} \rar C_{RR})~,\nnb \\
{\cal{V}}_{R_2} &=& {\cal{V}}_{L_2} ( C_{LL} \rar C_{LR}~,~~C_{RL} \rar C_{RR})~,\nnb \\
{\cal{V}}_{R_3} &=& {\cal{V}}_{L_3} ( C_{LL} \rar C_{LR}~,~~C_{RL} \rar C_{RR})~,\nnb \\
{\cal{V}}_{R_4} &=& {\cal{V}}_{L_4} ( C_{LL} \rar C_{LR}~,~~C_{RL} \rar C_{RR})~,\nnb \\
{\cal{S}}_{L} &=& - ( C_{LRRL} - C_{RLRL}) \ga \frac{2 m_{K^\ast}}{m_b} A_0 \dr~,\nnb \\
{\cal{S}}_{R} &=& - ( C_{LRLR} - C_{RLLR}) \ga \frac{2 m_{K^\ast}}{m_b} A_0 \dr~,\nnb 
\eea
where $q^{2}$ dependencies are implied. It is clear that all these effective form factors are functions 
of the specific form factors (\ref{ilk}--\ref{dort}) and the Wilson coefficients in (\ref{partoamp}). Therefore,
they carry information about both short-- and long--distance physics. 

The hadronic matrix element $\mbox{M}(B\rightarrow K^\ast \ell^{+}\ell^{-})$ is the basic machinery for the computation 
of all physical quantities pertaining to this decay. In particular, the computation of the energetic distributions,
total rate and various asymmetries follow from $(B\rightarrow K^\ast \ell^{+}\ell^{-})$ using the usual methods. In the 
next section, necessary asymmetries and other relevant quantities will be computed. 

\section{Asymmetries}

For an analysis of the asymmetries it is necessary to compute the differential decay rate for $B\rightarrow K^\ast
\ell^{+}\ell^{-}$ decay. For unpolarized leptons at the final state, using the decay amplitude 
in (\ref{hadroamp}), the differential decay rate is found to be 
\bea
\lefteqn{
\label{sifirrate}
\ga \frac{d \Gamma}{dq^2}\dr_0 = \frac{G^2 \alpha^2}{2^{14} \pi^5 m_B} 
\vel V_{tb} V_{ts}^\ast \ver^2 \lambda^{1/2} v} \nnb \\
&\times& \Bigg\{
32 \lambda m_B^4 \Big[ \frac{1}{3} (m_B^2 s - m_\ell^2)
(\vel  {\cal{V}}_{L_1} \ver^2 + \vel  {\cal{V}}_{R_1} \ver^2 ) + 2 m_\ell^2 \, \mbox{\rm Re} 
({\cal{V}}_{L_1}{\cal{V}}_{R_1}^\ast)\Big] \nnb \\
&+& 96 m_\ell^2 \, \mbox{\rm Re} ({\cal{V}}_{L_2} {\cal{V}}_{R_2}^\ast)-
\frac{4}{r} m_B^2 m_\ell \lambda \,
\mbox{\rm Re} [({\cal{V}}_{L_2}-{\cal{V}}_{R_2}) ({\cal{S}}_{L}^\ast-{\cal{S}}_{R}^\ast)] \nnb \\
&+&\frac{8}{r} m_B^2 m_\ell^2 \lambda \, 
\mbox{\rm Re} \Big[{\cal{V}}_{L_2}^\ast ({\cal{V}}_{L_4} +{\cal{V}}_{R_3}  - {\cal{V}}_{R_4})+
{\cal{V}}_{R_2}^\ast ({\cal{V}}_{L_3} - {\cal{V}}_{L_4} + {\cal{V}}_{R_4})-
({\cal{S}}_{L} {\cal{S}}_{R}^\ast) \Big] ~~~~~~~~ \nnb \\
&+&\frac{4}{r} m_B^4 m_\ell (1-r) \lambda \,
\Big\{\mbox{\rm Re} [({\cal{V}}_{L_3}-{\cal{V}}_{R_3}) ({\cal{S}}_{L}^\ast-{\cal{S}}_{R}^\ast)]\Big\} \nnb \\
&+& \frac{8}{r} m_B^4 m_\ell^2 (1-r) \lambda \,
\Big\{\mbox{\rm Re} [-({\cal{V}}_{L_3}-{\cal{V}}_{R_3}) ({\cal{V}}_{L_4}^\ast-{\cal{V}}_{R_4}^\ast)]\Big\}\nnb \\
&-& \frac{8}{r}m_B^4 m_\ell^2 \lambda (2+2 r-s)\, \mbox{\rm Re} ({\cal{V}}_{L_3} {\cal{V}}_{R_3}^\ast)
-\frac{4}{r} m_B^4 m_\ell s \lambda \,
\mbox{\rm Re} [({\cal{V}}_{L_4}-{\cal{V}}_{R_4})({\cal{S}}_{L}^\ast-{\cal{S}}_{R}^\ast)] \nnb \\
&-&\frac{4}{r} m_B^4 m_\ell^2 s \lambda \,
\Big[ \vel {\cal{V}}_{L_4} \ver^2 + \vel {\cal{V}}_{R_4} \ver^2 - 
2 \, \mbox{\rm Re}({\cal{V}}_{L_4} {\cal{V}}_{R_4}^\ast)\Big]
+\frac{2}{r} m_B^2 (m_B^2-2 m_\ell^2) \lambda \,
\Big[ \vel {\cal{S}}_{L} \ver^2 + \vel {\cal{S}}_{R} \ver^2 \Big] \nnb \\
&-&\frac{8}{3rs} m_B^2 \lambda \,
\Big[m_\ell^2 (2-2 r+s)+m_B^2 s (1-r-s) \Big]
\Big[\mbox{\rm Re}({\cal{V}}_{L_2}{\cal{V}}_{L_3}^\ast) + \mbox{\rm Re}({\cal{V}}_{R_2}{\cal{V}}_{R_3}^\ast)\Big] \nnb
\\ &+&\frac{4}{rs}\,
\Big[2 m_\ell^2 (\lambda-6 rs)+m_B^2 s (\lambda+12 rs) \Big]
\Big[ \vel {\cal{V}}_{L_2} \ver^2 + \vel {\cal{V}}_{R_2} \ver^2 \Big] \nnb \\
&+&\frac{4}{3rs} m_B^4 \lambda\,
\Big\{ m_B^2 s \lambda + m_\ell^2 \Big[ 2 \lambda + 3 s (2+2 r - s) \Big] \Big\}
\Big[ \vel {\cal{V}}_{L_3} \ver^2 + \vel {\cal{V}}_{R_3} \ver^2 \Big] 
\Bigg\}~.
\eea
where the subscript "0" is intended for the unpolarized decay rate. In this expression $s=q^{2}/m_{B}^{2}$,
$r=m_{K^\ast}^{2}/m_{B}^{2}$, $v^{2}=1-(4 m_{\ell}^2)/q^{2}$, and finally
$\lambda(1,r,s)=1+s^2+r^2-2r-2s-2rs$ is the familiar triangle function.

Our next task is the calculation of the lepton polarization asymmetries. For the 
computation of these asymmetries the unpolarized decay rate (\ref{sifirrate})
is not sufficient. The measurement of these asymmetries require the specification 
of the total number of leptons of a given kind (for example, negatively-charged)
in a given direction. Therefore, it  is necessary to take into account the 
polarization of the lepton beam in a given direction. Considering, for example, the 
negatively-charged lepton, one can introduce the following three polarization vectors
in the rest frame of $\ell^{-}$:
\bea
\vec{e}_L &=& \frac{\vec{p}_{-}}{\ve \vec{p}_{-} \ve}~, \nnb \\
\vec{e}_N &=& \frac{\vec{p}_{K^\ast}\times\vec{p}_{-}}
                   {\ve \vec{p}_{K^\ast}\times\vec{p}_{-} \ve}~, \nnb \\
\vec{e}_T &=&
\frac{(\vec{p}_{K^\ast}\times\vec{p}_{-})\times\vec{p}_{-}}{\ve
(\vec{p}_{K^\ast}\times\vec{p}_{-})\times\vec{p}_{-}\ve}~,
\eea
where $\vec{e}_{i}\cdot \vec{e}_{j}=\delta_{i,j}$, $p_{-}\cdot \vec{e}_{i}=0$,
$i,j=L,T,N$. Here, $\vec{e}_L$, $\vec{e}_T$ and $\vec{e}_N$ correspond, 
respectively, to the `longitudinal', `transversal' and `normal'
polarization directions of $\ell^{-}$ with respect to its direction of motion, $\vec{e}_L$.
One notices that $\vec{e}_L$ and $\vec{e}_T$ are co--planar, and $\vec{e}_N$ is perpendicular 
to this plane. In the rest frame of $\ell^{-}$, the temporal components of the corresponding 
four--vectors vanish. However, in the  dilepton rest frame (that is, $\vec{q}=0$), 
the four--vector corresponding to $\vec{e}_L$ is boosted to 
$(\vec{p}_{-}/m_{\ell}, (E_{\ell}/m_{\ell})\vec{e}_L)$ leaving $\vec{e}_T$ and $\vec{e}_N$
unchanged. In the following, all results will be conveniently given in the dilepton rest frame.

The differential decay rate for any spin direction $\vec{n}$
of the $\ell^-$, where $\vec{n}$ is a unit vector in the $\ell^-$
rest frame satisfying $\vec{n}\cdot\vec{n}=1, \vec{n}\cdot \vec{p}_{-}=0$, 
can be expressed in the following form
\bea
\label{acil}
\frac{d \Gamma(\vec{n})}{dq^2} = \frac{1}{2} \ga \frac{d \Gamma}{dq^2}\dr_0
\Big[1 + \Big( P_L \vec{e}_L + P_N \vec{e}_N + P_T \vec{e}_T \Big) \cdot
\vec{n}\Big]~,
\eea
where the coefficients of unit vectors, $P_L$, $P_N$ and $P_T$,
are recognized as the `longitudinal', `normal' and `transversal' polarization
asymmetries. A simple formula for extracting these asymmetries from the polarized
decay rate follows from (\ref{acil}) itself:
\bea
\label{asym}
P_i (q^2) = \frac{\ds{\frac{d \Gamma}{dq^2} (\vec{n}=\vec{e}_i) -
                      \frac{d \Gamma}{dq^2} (\vec{n}=-\vec{e}_i)}} 
                 {\ds{\frac{d \Gamma}{dq^2} (\vec{n}=\vec{e}_i) +
                      \frac{d \Gamma}{dq^2} (\vec{n}=-\vec{e}_i)}}~,
\eea
where $i=L,T,N$. One notes that the denominator in this expression is identical to the 
unpolarized decay rate (\ref{sifirrate}). On the other hand, the numerator depends on the 
spin direction of the lepton under consideration. In essence what $P_i (q^2)$ measures is 
the difference between the rates for a particular direction and its opposite for a given 
dilepton invariant mass $m_{\ell\ell}=\sqrt{q^{2}}$.

Using the hadronic decay amplitude (\ref{hadroamp}) in the general polarization asymmetry formulae (\ref{asym}),
after a lengthy calculation the polarization asymmetries $P_L$, $P_N$ and $P_T$
are found to have the following explicit expressions:
\bea
\label{polasym}
P_L &=& \frac{1}{\Delta} v \Bigg\{ 
\frac{4}{3 r} \lambda^2 m_B^6 \Big[ \vel {\cal{V}}_{L_3} \ver^2 - \vel {\cal{V}}_{R_3} \ver^2\Big] +
\frac{4}{r} \lambda m_B^2 m_\ell \, 
\mbox{\rm Re} [({\cal{V}}_{L_2} - {\cal{V}}_{R_2}) ({\cal{S}}_{L}^\ast + {\cal{S}}_{R}^\ast)] \nnb \\
&-& \frac{4}{r} \lambda m_B^4 m_\ell (1-r) \, 
\mbox{\rm Re} [({\cal{V}}_{L_3} - {\cal{V}}_{R_3}) ({\cal{S}}_{L}^\ast + {\cal{S}}_{R}^\ast)]+
\frac{32}{3} \lambda m_B^6 s \Big[ \vel {\cal{V}}_{L_1} \ver^2 - \vel {\cal{V}}_{R_1} \ver^2\Big] \nnb \\
&-&\frac{2}{r} \lambda m_B^4 s 
\Big[ \vel {\cal{S}}_{L} \ver^2 - \vel {\cal{S}}_{R} \ver^2\Big]+
\frac{4}{r} \lambda m_B^4 m_\ell s \, 
\mbox{\rm Re} [({\cal{V}}_{L_4} - {\cal{V}}_{R_4}) ({\cal{S}}_{L}^\ast + {\cal{S}}_{R}^\ast)] \nnb \\
&-&\frac{8}{3 r} \lambda m_B^4 (1-r-s)
\Big[ \mbox{\rm Re}({\cal{V}}_{L_2}{\cal{V}}_{L_3}^\ast) - \mbox{\rm Re}({\cal{V}}_{R_2}{\cal{V}}_{R_3}^\ast)\Big]\nnb
\\
&+&\frac{4}{3 r} \lambda m_B^2 (\lambda + 12 r s) 
\Big[ \vel {\cal{V}}_{L_2} \ver^2 - \vel {\cal{V}}_{R_2} \ver^2\Big] \Bigg\}~, \nnb \\ \nnb \\
P_T &=& \frac{1}{\Delta} \sqrt{\lambda} \pi \Bigg\{ 
-8 m_B^3 m_\ell \sqrt{s} \, \mbox{\rm Re} [({\cal{V}}_{L_1} + {\cal{V}}_{R_1}) ({\cal{V}}_{L_2}^\ast +
{\cal{V}}_{R_2}^\ast)] \nnb \\
&+& \frac{1}{r} m_B^3 m_\ell (1+3 r + s) \sqrt{s} \, 
\Big[ \mbox{\rm Re}({\cal{V}}_{L_2}{\cal{V}}_{R_3}^\ast) - \mbox{\rm Re}({\cal{V}}_{L_3}{\cal{V}}_{R_2}^\ast)\Big]
\nnb \\
&+&\frac{1}{r\sqrt{s}} m_B m_\ell (1- r - s)
\Big[ \vel {\cal{V}}_{L_2} \ver^2 - \vel {\cal{V}}_{R_2} \ver^2\Big] \nnb \\
&+&\frac{2}{r\sqrt{s}} m_B m_\ell^2 (1- r - s)
\Big[ \mbox{\rm Re}({\cal{V}}_{L_2} {\cal{S}}_{R}^\ast) - \mbox{\rm Re}({\cal{V}}_{R_2} {\cal{S}}_{L}^\ast)\Big]
\nnb \\
&+&\frac{1}{r} m_B^3 m_\ell (1- r - s) \sqrt{s}\,
\mbox{\rm Re} [({\cal{V}}_{L_2} + {\cal{V}}_{R_2}) ({\cal{V}}_{L_4} - {\cal{V}}_{R_4})] \nnb \\
&+&\frac{2}{r\sqrt{s}} m_B^3 m_\ell^2 \lambda 
\Big[ -\mbox{\rm Re}({\cal{V}}_{L_3} {\cal{S}}_{R}^\ast) + \mbox{\rm Re}({\cal{V}}_{R_3} {\cal{S}}_{L}^\ast)\Big]
\nnb \\ 
&+&\frac{1}{r\sqrt{s}} m_B^5 m_\ell(1-r) \lambda
\Big[ \vel {\cal{V}}_{L_3} \ver^2 - \vel {\cal{V}}_{R_3} \ver^2\Big]\nnb \\
&+& \frac{1}{r} m_B^5 m_\ell \lambda \sqrt{s} \,
\mbox{\rm Re} [-({\cal{V}}_{L_3} + {\cal{V}}_{R_3}) ({\cal{V}}_{L_4}^\ast - {\cal{V}}_{R_4}^\ast)] \nnb \\
&+&\frac{1}{r\sqrt{s}} m_B^3 m_\ell [(1-r-s) ( 1-r) + \lambda ]
\Big[ -\mbox{\rm Re}({\cal{V}}_{L_2}{\cal{V}}_{L_3}^\ast) + \mbox{\rm Re}({\cal{V}}_{R_2}{\cal{V}}_{R_3}^\ast)\Big]
\nnb \\
&+&\frac{1}{r\sqrt{s}} m_B (1-r-s)(-2 m_\ell^2  + m_B^2 s )
\Big[\mbox{\rm Re}({\cal{V}}_{R_2}{\cal{S}}_{R}^\ast) - \mbox{\rm Re}({\cal{V}}_{L_2} {\cal{S}}_{L}^\ast)\Big] \nnb
\\
&+&\frac{1}{r\sqrt{s}} m_B^3 \lambda (-2 m_\ell^2  + m_B^2 s )
\Big[-\mbox{\rm Re}({\cal{V}}_{R_3}{\cal{S}}_{R}^\ast) + \mbox{\rm Re}({\cal{V}}_{L_3} {\cal{S}}_{L}^\ast)\Big]
 \Bigg\}~, \nnb \\ \nnb \\ 
P_N &=& \frac{1}{\Delta} \pi v m_B^3 \sqrt{\lambda} \sqrt{s} \Bigg\{
8 m_\ell \, \mbox{\rm Im}({\cal{V}}_{L_2}^\ast {\cal{V}}_{R_1} + {\cal{V}}_{L_1}^\ast {\cal{V}}_{R_2}) \nnb \\
&+& \frac{1}{r} m_\ell (1+3 r-s)\,
\mbox{\rm Im} \Big[({\cal{V}}_{L_2}+{\cal{V}}_{R_2}) ({\cal{V}}_{L_3}^\ast - {\cal{V}}_{R_3}^\ast)\Big] \nnb \\
&+&\frac{1}{r} m_B^2 \lambda 
\mbox{\rm Im}\Big[(m_\ell {\cal{V}}_{L_4}-m_\ell {\cal{V}}_{R_4} - {\cal{S}}_{L}) {\cal{V}}_{L_3}^\ast -
(m_{\ell}{\cal{V}}_{R_4}-m_\ell {\cal{V}}_{L_4}- {\cal{S}}_{R}) {\cal{V}}_{R_3}^\ast \Big] \nnb \\
&+& \frac{1}{r} (1-r-s) 
\mbox{\rm Im}\Big[({\cal{S}}_{L} - m_\ell {\cal{V}}_{L_4}+m_\ell {\cal{V}}_{R_4}) {\cal{V}}_{L_2}^\ast -
({\cal{S}}_{R}-m_\ell {\cal{V}}_{R_4} + m_\ell {\cal{V}}_{L_4}) {\cal{V}}_{R_2}^\ast \Big] \Bigg\}~,
\eea
where $\Delta$ is the expression within the curly parenthesis in the
unpolarized differential decay rate in (\ref{sifirrate}). These expressions
for the polarization asymmetries are quite general except for the neglect of strange
quark mass and the tensor operators (the last two operators in (\ref{partoamp})) as mentioned before. 

Before proceeding, it is convenient to make a few useful observations on the lepton asymmetries. Particularly 
interesting one is the massless (light) lepton limit: $m_{\ell} \rightarrow 0$. In this case, $P_L$
depends only 
on the bilinears of ${\cal{V}}_{L_i}$ and ${\cal{V}}_{R_i}$, that is, the effects of the scalar operators in 
(\ref{partoamp}) completely decouple. On the other hand, $P_T$ and $P_N$ happen to depend only on the interference terms
between the coefficients of the vector operators (${\cal{V}}_{L_i}$ and ${\cal{V}}_{R_i}$) and those of the 
scalar operators (${\cal{S}}_{L,R}$). However, one notices that the scalar operators in (\ref{partoamp}) can be induced
by an exchange of the scalar particle (such as two Higgs doublet models \cite{erhanabi}) in which case the coefficients
${\cal{S}}_{L,R}$ are necessarily proportional to the lepton mass. Therefore, in the limit of massless (light) leptons 
only the longitudinal asymmetry $P_L$ can remain non--vanishing. Conversely, in near future, if 
experiment
yields non--vanishing $P_N$ and $P_T$ for $B\rightarrow K^\ast e^+ e^-$ decay this would imply the
generation of scalar operators by mechanisms beyond the Higgs model where the fermion scalar coupling
is always proportional to the fermion mass.

In general, the lepton polarization asymmetries (\ref{polasym}) are able to probe real as well 
as imaginary parts of the effective form factors ${\cal{V}}_{L_i}$, ${\cal{V}}_{R_i}$ and ${\cal{S}}_{L,R}$.
As the parametrization  (\ref{formfac}) shows the hadronic form factors are inherently real, and thus the 
imaginary parts of ${\cal{V}}_{L_i}$, ${\cal{V}}_{R_i}$ and ${\cal{S}}_{L,R}$ in (\ref{hadroamp}) can come only from the Wilson
coefficients in (\ref{partoamp}). Below we will keep this picture, that is, we will be dealing only with the CP violation 
effects due to short--distance physics parametrized by the Wilson coefficients. 
At this point it is useful to distinguish between CP properties of the phases in the Wilson coefficients. 
In principle, $C_{BR},\cdots C_{TE}$ all can have finite phases; however, these phases can have 
$strong$ and  $weak$ subparts. Here by $strong$ and  $weak$ we mean $even$ and $odd$ phases under 
CP conjugation. To be able to distinguish such distinct components of the phases it is not sufficient to analyze
the polarization asymmetries alone. One, in particular, has to consider the CP asymmetry of the decay 
which is inherently sensitive to CP character of the phases of the Wilson coefficients. Using the unpolarized decay 
rate (\ref{sifirrate}), the CP asymmetry for  $B\rightarrow K^{\ast} \ell^{+}\ell^{-}$ decay is defined by:
\bea
\label{CPasym}
A_{CP} (q^2) = \frac{\ds{\left(\frac{d \Gamma}{dq^2}\right)_{0} (B\rightarrow K^\ast \ell^{+}\ell^{-}) -
                       \left(\frac{d \Gamma}{dq^2}\right)_{0}(\overline{B}\rightarrow \overline{K^\ast}
\ell^{+}\ell^{-})}}
                 {\ds{\left(\frac{d \Gamma}{dq^2}\right)_{0} (B\rightarrow K^\ast \ell^{+}\ell^{-}) +
                      \left(\frac{d \Gamma}{dq^2}\right)_{0} (\overline{B}\rightarrow \overline{K^\ast}
\ell^{+}\ell^{-})}}~,
\eea 
where the processes to which  $d \Gamma/dq^2$ refers are explicitly shown in the 
arguments. Making use of the explicit expression for the unpolarized decay rate (\ref{sifirrate})
one can determine the detailed dependence of $A_{CP} (q^2)$ on the model parameters.
For this purpose it is useful to introduce the following parametrization for the quantities
${\cal{V}}_{L_i}$, ${\cal{V}}_{R_i}$ and ${\cal{S}}_{L,R}$ (\ref{hadroamp}):
\begin{eqnarray}
\label{phasedefine}
{\cal{V}}_{L_i}=|{\cal{V}}_{L_i}|e^{i\phi_{w}^{L_i}+i\phi_{s}^{L_i}}~,\ \
{\cal{V}}_{R_i}=|{\cal{V}}_{R_i}|e^{i\phi_{w}^{R_i}+i\phi_{s}^{R_i}}~, \ \
{\cal{S}}_{L,R}=|{\cal{V}}_{L,R}|e^{i\phi_{w}^{L,R}+i\phi_{s}^{L,R}}~,
\end{eqnarray} 
where $i=1,\cdots,4$. In this expression subscript "$s$" ("$w$") stands for $strong$ ($weak$) phases mentioned above. 
By definition, ${\cal{V}}$'s and ${\cal{S}}$'s are combinations of hadronic form factors and Wilson coefficients so that
the phases $\phi_{w,s}$ defined by (\ref{phasedefine}) are explicit functions of the dilepton invariant mass. With this definition
of the from factors it is possible to find a suggestive form for the CP asymmetry:
\bea
\label{acp}
\lefteqn{
A_{CP}(q^2) = \frac{1}{\Sigma} \Bigg\{
- 64 \lambda m_B^4 m_\ell^2 \, \vel {\cal{V}}_{L_1} \ver \vel {\cal{V}}_{R_1} \ver \sin\Delta\phi_{s}^{L_1, R_1}
\sin \Delta\phi_{w}^{L_1, R_1}}\nnb \\ &-& 96 m_\ell^2 \vel {\cal{V}}_{L_2} \ver \vel {\cal{V}}_{R_2}\ver
\sin\Delta\phi_{s}^{L_2, R_2}
\sin \Delta\phi_{w}^{L_2, R_2}+\frac{4}{r} m_B^2 m_\ell \lambda \,\Big[\vel {\cal{V}}_{L_2} \ver \vel
{\cal{S}}_{L}\ver
\sin\Delta\phi_{s}^{L_2, L} \sin \Delta\phi_{w}^{L_2, L}\nnb \\ &+&\vel {\cal{V}}_{R_2} \ver \vel {\cal{S}}_{R}\ver
\sin\Delta\phi_{s}^{R_2, R} \sin \Delta\phi_{w}^{R_2, R}-\vel {\cal{V}}_{L_2} \ver \vel {\cal{S}}_{R}\ver
\sin\Delta\phi_{s}^{L_2, R} \sin \Delta\phi_{w}^{L_2, R}\nnb \\ &-&\vel {\cal{V}}_{R_2} \ver \vel {\cal{S}}_{L}\ver
\sin\Delta\phi_{s}^{R_2, L} \sin \Delta\phi_{w}^{R_2, L}\Big ]-
\frac{4}{r} m_B^4 m_\ell (1-r) \lambda \,\Big[\vel {\cal{V}}_{L_3} \ver \vel {\cal{S}}_{L}\ver
\sin\Delta\phi_{s}^{L_3, L} \sin \Delta\phi_{w}^{L_3, L}\nnb \\ &+&\vel {\cal{V}}_{R_3} \ver \vel {\cal{S}}_{R}\ver
\sin\Delta\phi_{s}^{R_3, R} \sin \Delta\phi_{w}^{R_3, R}-\vel {\cal{V}}_{L_3} \ver \vel {\cal{S}}_{R}\ver
\sin\Delta\phi_{s}^{L_3, R} \sin \Delta\phi_{w}^{L_3, R}\nnb \\ &-&\vel {\cal{V}}_{R_3} \ver \vel {\cal{S}}_{L}\ver
\sin\Delta\phi_{s}^{R_3, L} \sin \Delta\phi_{w}^{R_3, L}\Big]+
\frac{4}{r} m_B^4 m_\ell s \lambda \,\Big[\vel {\cal{V}}_{L_4} \ver \vel {\cal{S}}_{L}\ver
\sin\Delta\phi_{s}^{L_4, L} \sin \Delta\phi_{w}^{L_4, L}\nnb \\ &+&\vel {\cal{V}}_{R_4} \ver \vel {\cal{S}}_{R}\ver
\sin\Delta\phi_{s}^{R_4, R} \sin \Delta\phi_{w}^{R_4, R}-\vel {\cal{V}}_{L_4} \ver \vel {\cal{S}}_{R}\ver
\sin\Delta\phi_{s}^{L_4, R} \sin \Delta\phi_{w}^{L_4, R}\nnb \\ &-&\vel {\cal{V}}_{R_4} \ver \vel {\cal{S}}_{L}\ver
\sin\Delta\phi_{s}^{R_4, L} \sin \Delta\phi_{w}^{R_4, L}\Big]+
\frac{8}{r} m_B^4 m_\ell^2 (1-r) \lambda \,\Big[\vel {\cal{V}}_{L_3} \ver \vel {\cal{V}}_{L_4}\ver
\sin\Delta\phi_{s}^{L_3, L_4} \sin \Delta\phi_{w}^{L_3, L_4}\nnb \\&+&\vel {\cal{V}}_{R_3} \ver \vel {\cal{V}}_{R_4}\ver
\sin\Delta\phi_{s}^{R_3, R_4} \sin \Delta\phi_{w}^{R_3, R_4}-\vel {\cal{V}}_{L_3} \ver \vel {\cal{V}}_{R_4}\ver
\sin\Delta\phi_{s}^{L_3, R_4} \sin \Delta\phi_{w}^{L_3, R_4}\nnb \\ &-&\vel {\cal{V}}_{R_3} \ver \vel {\cal{V}}_{L_4}\ver
\sin\Delta\phi_{s}^{R_3, L_4} \sin \Delta\phi_{w}^{R_3, L_4}\Big]\nnb \\ &+&
\frac{8}{r}m_B^4 m_\ell^2 \lambda (2+2 r-s)\,\vel {\cal{V}}_{L_3} \ver \vel {\cal{V}}_{R_3}\ver \sin\Delta\phi_{s}^{L_3,
R_3}\sin \Delta\phi_{w}^{L_3, R_3}\nnb \\ &-&
\frac{8}{r} m_B^4 m_\ell^2 s \lambda \,\vel {\cal{V}}_{L_4} \ver \vel {\cal{V}}_{R_4}\ver \sin\Delta\phi_{s}^{L_4,
R_4}\sin \Delta\phi_{w}^{L_4, R_4}\nnb \\&+&
\frac{8}{3rs} m_B^2 \lambda \,\Big[\vel {\cal{V}}_{L_2} \ver \vel {\cal{V}}_{L_3}\ver \sin\Delta\phi_{s}^{L_2,
L_3}\sin \Delta\phi_{w}^{L_2, L_3}+\vel {\cal{V}}_{R_2} \ver \vel {\cal{V}}_{R_3}\ver \sin\Delta\phi_{s}^{R_2,
R_3}\sin \Delta\phi_{w}^{R_2, R_3}\Big]\nnb \\&-&
\frac{8}{r}m_B^2 m_\ell^2 \lambda \,\Big[\vel {\cal{V}}_{L_2} \ver \vel {\cal{V}}_{L_4}\ver \sin\Delta\phi_{s}^{L_2,
L_4}\sin \Delta\phi_{w}^{L_2, L_4}+\vel {\cal{V}}_{L_2} \ver \vel {\cal{V}}_{R_3}\ver \sin\Delta\phi_{s}^{L_2,
R_3}\sin \Delta\phi_{w}^{L_2, R_3}\nnb \\ &-&\vel {\cal{V}}_{L_2} \ver \vel {\cal{V}}_{R_4}\ver \sin\Delta\phi_{s}^{L_2,
R_4}\sin \Delta\phi_{w}^{L_2, R_4}+\vel {\cal{V}}_{R_2} \ver \vel {\cal{V}}_{R_4}\ver \sin\Delta\phi_{s}^{R_2,
R_4}\sin \Delta\phi_{w}^{R_2, R_4}\nnb \\ &+&\vel {\cal{V}}_{R_2} \ver \vel {\cal{V}}_{L_3}\ver \sin\Delta\phi_{s}^{R_2,
L_3}\sin \Delta\phi_{w}^{R_2, L_3}-\vel {\cal{V}}_{R_2} \ver \vel {\cal{V}}_{L_4}\ver \sin\Delta\phi_{s}^{R_2,
L_4}\sin \Delta\phi_{w}^{R_2, L_4}\nnb\\&-&\vel {\cal{S}}_{L} \ver \vel {\cal{S}}_{R}\ver \sin\Delta\phi_{s}^{L,
R}\sin \Delta\phi_{w}^{L,R}\Big]\Bigg\}~,
\eea
where $\Delta\phi_{x}^{a,b}\equiv\phi_{x}^{a}-\phi_{x}^{b}$. The quantity $\Sigma$ in the denominator is even under all these
phases, and has the expression 
\bea
\Sigma &=& \mbox{Numerator of $A_{CP}$}\Big(\sin \Delta\phi_{s}^{a,b} \sin \Delta\phi_{w}^{a,b} \longrightarrow - \cos
\Delta\phi_{s}^{a,b} \cos \Delta\phi_{w}^{a,b} \Big)\nnb \\ &+&\Bigg\{
\frac{32}{3} \lambda m_B^4 (m_B^2 s - m_\ell^2)
\Big[\vel  {\cal{V}}_{L_1} \ver^2 + \vel  {\cal{V}}_{R_1} \ver^2 \Big]-\frac{4}{r} m_B^4 m_\ell^2 s \lambda \,
\Big[\vel {\cal{V}}_{L_4} \ver^2 + \vel {\cal{V}}_{R_4} \ver^2\Big]\nnb \\&+&
\frac{2}{r} m_B^2 (m_B^2-2 m_\ell^2) \lambda \,
\Big[\vel {\cal{S}}_{L} \ver^2 + \vel {\cal{S}}_{R} \ver^2 \Big]\nnb\\&+&\frac{4}{rs}\,
\Big[2 m_\ell^2 (\lambda-6 rs)+m_B^2 s (\lambda+12 rs) \Big]
\Big[ \vel {\cal{V}}_{L_2} \ver^2 + \vel {\cal{V}}_{R_2} \ver^2 \Big] \nnb \\
&+&\frac{4}{3rs} m_B^4 \lambda\,
\Big\{ m_B^2 s \lambda + m_\ell^2 \Big[ 2 \lambda + 3 s (2+2 r - s) \Big] \Big\}
\Big[ \vel {\cal{V}}_{L_3} \ver^2 + \vel {\cal{V}}_{R_3} \ver^2 \Big]
\Bigg\}~.
\eea

Until now the decay rate (\ref{sifirrate}), the lepton polarization asymmetries (\ref{polasym})
and CP asymmetry (\ref{CPasym}) have been computed by adopting a rather general quark level 
amplitude (\ref{partoamp}) for $B\rightarrow K^{\ast} \ell^{+}\ell^{-}$ decay. Presently this 
exclusive decay has a direct bound coming from recent CDF measurement \cite{cdf1}: BR($B \rightarrow
K^\ast \mu^+\mu^-)< 4. 0 \times 10^{-6}$. In addition to this direct bound, existing CLEO
result \cite{CLEO} for BR($B\rightarrow K^{*} \gamma$) imposes another important, albeit partial,
constraint on the parameter space. Indeed, using the notation of (\ref{partoamp}) and appropriate form
factors derived in (\ref{ilk})-(\ref{dort}) the total decay rate for 
$B\rightarrow K^\ast \gamma$ can be written as
\begin{eqnarray}
\label{bkgam}
\Gamma(B\rightarrow K^\ast \gamma)=\frac{G^2 \alpha m_B^{3} m_{b}^{2}}{128 \pi^4} \vel V_{ts} V_{tb}^* \ver^2
\left( 1- \frac{m_{K}^{2}}{m_{B}^{2}}\right)^{3} \vel C_{BR}  T_{1}(0) \ver^{2}~,
\end{eqnarray}
which constrains directly $|C_{BR}~T_{1}(0)|$. Therefore, the norm of the Wilson coefficient for the
dipole operator is 
fixed by BR($B\rightarrow K^{*} \gamma$). This constraint, however, does not say anything about the CP violation
potential of $C_{BR}$. 

\section{Large CP--phases within an SM--like model}
As as been emphasized before, the underlying model for the partonic decay amplitude
(\ref{partoamp}) is rather general one. In principle, all the Wilson coefficients can be non-zero 
and may have arbitrary phases. However, a realistic model should meet the requirements of the 
SM to leading order, and especially, should not cause unacceptable deviations from the existing 
experimental data confirmed already by the SM. For this reason it is convenient 
to establish the discussion on a general model with close reference to the SM predictions. 

Therefore, here we follow a minimal prescription such that the general Wilson coefficients in (\ref{partoamp}) are
\begin{itemize}
\item endowed with new phases beyond the SM, 
\item identical to SM ones when these phases vanish.
\end{itemize}
Such an approach obviously neglects the new physics contributions to the norms of the Wilson coefficients; however, at 
the aim of determining the information content of the polarization and CP asymmetries on the sources of CP violation, it 
suffices. This is kind of a minimal approach for parametrizing the new physics CP violation for asymmetry measurements in 
$B\rightarrow K^{\ast} \ell^{+}\ell^{-}$ decay.  
 
Adopting this approach, one can make the following assignments for the general Wilson coefficients in (\ref{partoamp}).
First, the coefficients describing the scalar--scalar type interactions vanish identically 
\bea
C_{LRRL} = C_{RLLR} = C_{LRLR} = C_{RLRL} = 0.
\eea
It is known that such coefficients exist, for example, in the two Higgs doublet models (2HDM) which have an extended Higgs
sector compared to the SM. In such models these scalar--scalar interactions are induced by the Higgs exchange, and the 
resulting Wilson coefficients are proportional to $m_{b}m_{\ell}/m_{h}^{2}$ which is maximal for $\ell=\tau$. However, to the 
extent one neglects $m_s/m_b$, $m_{b}m_{\ell}/m_{h}^{2}$ is, too, negligible in the light of recent LEP
limit on the Higgs boson mass $m_h$ \cite{LEP} is taken into account.    

The Wilson coefficient for the dipole operator $C_{BR}$ obeys 
\begin{eqnarray}
\label{cbr}
C_{BR}=-2 C_{7}^{eff}(m_b)~e^{i \phi_{7}}\equiv -2\left( C_{7}(m_{b})-\frac{1}{3} C_{5}(m_{b})-C_{6}(m_{b})\right)~e^{i \phi_{7}}
\end{eqnarray}
where $\phi_7$ is an arbitrary phase, and it is not constrained by BR($B\rightarrow K^{*} \gamma$) at all.

Finally the coefficients of the vector--vector interactions in (\ref{partoamp}) are given by 
\begin{eqnarray}
C_{LL (LR)}=C_{9}^{eff}(m_{b})~e^{i \phi_{9}}- (+) C_{10}(m_{b})~e^{i \phi_{10}}~,\ \ C_{RL}=C_{RR}=0~
\end{eqnarray}
where the coefficient $C_{10}$ is known to be scale independent: $C_{10}(m_{b})= C_{10}(M_{W})$. 

In the SM the Wilson coefficients $C_7^{eff}(m_b)$ and $C_{10}(m_b)$ are strictly real as can be read off from Table II. Moreover, the
SM prediction for $C_7^{eff}(m_b)$ is already consistent with the CLEO determination of BR($B\rightarrow K^{*} \gamma$) in
(\ref{bkgam}). Therefore, through the choice of $C_{BR}$ in (\ref{cbr}) the experimental constraint is already taken into account.
Although individual Wilson coefficients at $\mu \sim m_b$ level are all real (see Table II) the effective Wilson coefficient
$C_{9}^{eff}(m_{b}, q^{2})$ has a finite phase, and is an explicit function of the dilepton invariant mass, $q^2$. To see its phase
content it is useful to reproduce its explicit expression here:
\begin{eqnarray}
\label{c9}
C_{9}^{eff}(m_{b})=C_{9}(m_{b})\Bigg\{ 1 + \frac{\alpha_s \ga \mu \dr }{\pi}
\omega \ga \hat s \dr \Bigg\} + Y_{SD}\ga m_{b}, \hat s \dr+Y_{LD} \ga m_{b}, \hat s \dr~
\end{eqnarray}
where $C_{9}(m_{b})$ is read off from Table II. Here $\omega \ga \hat s \dr$ represents the ${\cal{O}}(\alpha_{s})$ corrections
coming from one--gluon exchange in the matrix element of the corresponding operator ${\cal{O}}_{9}$ \cite{R17}: 
\bea
\omega \ga \hat s \dr &=& - \frac{2}{9} \pi^2 -
\frac{4}{3} Li_2  \ga \hat s \dr - \frac{2}{3} \ln \ga \hat s\dr
\,\ln \ga 1 -\hat s \dr
- \,\frac{5 + 4 \hat s}{3 \ga 1 + 2 \hat s \dr} \ln \ga 1 -\hat s \dr \nnb \\
&-&\frac{2 \hat s \ga 1 + \hat s \dr \ga 1 - 2 \hat s \dr}
{3 \ga 1 - \hat s \dr^2 \ga 1 + 2 \hat s \dr} \, \ln \ga \hat s\dr
+ \frac{5 + 9 \hat s - 6 {\hat s}^2}
{3 \ga 1 - \hat s \dr \ga 1 + 2 \hat s \dr}~.
\eea

In (\ref{c9}) $Y_{SD}$ and $Y_{LD}$ represent, respectively, the short-- and long--distance contributions of 
the four--quark operators ${\cal{O}}_{i=1,\cdots,6}$ \cite{R17,R18}. Here $Y_{SD}$ can be obtained
by a perturbative calculation  
\bea
Y_{SD}\ga m_{b}, \hat s \dr &=& g \ga \hat m_c,\hat s \dr
\left[3 C_1 + C_2 + 3 C_3 + C_4 + 3 C_5 + C_6 \right] \nnb \\
&-& \frac{1}{2} g \ga 1,\hat s \dr
\left[4 C_3 +4 C_4 + 3 C_5 + C_6 \right] \nnb \\
&-& \frac{1}{2} g \ga 0,\hat s \dr
\left[ C_3 + 3  C_4 \right]
+ \frac{2}{9} \left[ 3 C_3 + C_4 + 3 C_5 + C_6 \right] \nnb \\
&-& \frac{V_{us}^* V_{ub}}{V_{ts}^* V_{tb}}
\left[ 3 C_1 + C_2 \right] \left[ g \ga 0,\hat s \dr -
g \ga \hat m_c,\hat s \dr \right]~,
\eea
where the loop function $g \ga m_q, s \dr$ stands for the loops
of quarks with mass $m_{q}$ at the dilepton invariant mass $s$.
This function develops absorbtive parts for dilepton energies 
$s= 4 m_q^{2}$:
\bea
\lefteqn{
g \ga \hat m_q,\hat s \dr = - \frac{8}{9} {\rm ln} \hat m_q +
\frac{8}{27} + \frac{4}{9} y_q -
\frac{2}{9} \ga 2 + y_q \dr \sqrt{\vel 1 - y_q \ver}} \nnb \\   
&&\times \Big\{ \Theta(1 - y_q)
\ga {\rm ln} \frac{1  + \sqrt{1 - y_q}}{1  -  \sqrt{1 - y_q}} - i \pi \dr
+ \Theta(y_q - 1) 2 \,{\rm arctan} \frac{1}{\sqrt{y_q - 1}} \Big\},
\eea
where  $\hat m_q= m_{q}/m_{b}$ and $y_q=4 \hat m_q^2/\hat s$. Hence, due to 
the absorbtive parts of $g \ga \hat m_q,\hat s \dr$, there are strong phases
coming from $Y_{SD}$. One, in particular, notices the terms proportional 
to $g \ga 0,\hat s \dr$ which have a non--vanishing imaginary parts independent
of the dilepton invariant mass. 

In addition to these perturbative contributions, the $\bar{c}c$ loops
can excite low--lying charmonium states $\psi(1s), \cdots, \psi(6s)$
whose contributions are represented by $Y_{LD}$ \cite{bw}: 
\bea
Y_{LD}\ga m_{b}, \hat s \dr&=& \frac{3}{\alpha^2} 
\Bigg\{ - \frac{V_{cs}^* V_{cb}}{V_{ts}^* V_{tb}} \,C^{\ga 0 \dr} -
\frac{V_{us}^* V_{ub}}{V_{ts}^* V_{tb}}
\left[ 3 C_3 + C_4 + 3 C_5 + C_6 \right] \Bigg\} \nnb \\
&\times& \sum_{V_i = \psi \ga 1 s \dr, \cdots, \psi \ga 6 s \dr}
\ds{\frac{ \pi \kappa_{i} \Gamma \ga V_i \rar \ell^+ \ell^- \dr M_{V_i} }
{\ga M_{V_i}^2 - \hat s m_b^2 - i M_{V_i} \Gamma_{V_i} \dr }}~,
\eea
where $\kappa_i$ are the Fudge factors for $B\rightarrow K^\ast V_{i} \rightarrow K^\ast \ell^{+}\ell^{-}$ transition, and
$C^{\ga 0 \dr} \equiv 3 C_1+C_2 + 3 C_3 + C_4 + 3 C_5 + C_6$. Here the sum runs over all 
all charmonium resonances with mass $m_{V_{i}}$ and total decay rate $\Gamma_{V_{i}}$. Contrary to $Y_{SD}$, the 
long--distance contribution $Y_{LD}$ has both weak and strong phases. The weak phases follow from the 
CKM elements whereas the strong phases come from the $\hat s$ values for which $i$-th charmonium state is on shell.
Therefore, the Wilson coefficient $C_{9}^{eff}(m_{b})$ has both weak and strong phases already in the SM.

\begin{table}[ht]
\renewcommand{\arraystretch}{1.5}
\addtolength{\arraycolsep}{3pt}
$$
\begin{array}{|l|l|l|l|l|l|l|l|l|}
\hline
C_{1}(m_b) & C_{2}(m_b) & C_{3}(m_b) & C_{4}(m_b) & C_{5}(m_b) & C_{6}(m_b) & C_{7}^{eff}(m_b) &
C_{9}(m_b) & C_{10}(m_b)\\ \hline
-0.248 & 1.107& 0.011& -0.026& 0.007& -0.031& -0.313& 4.344& -4.669\\ \hline
\end{array}
$$
\caption{The numerical values of the Wilson coefficients at $\mu\sim m_{b}$ scale within the SM. The corresponding
numerical value of $C^{0}$ is 0.362.}
\renewcommand{\arraystretch}{1}
\addtolength{\arraycolsep}{-3pt}
\end{table}

In this sense, the Wilson coefficients $C_7^{eff}(m_b)$ and $C_{10}(m_b)$ cannot develop any strong phase, 
and thus, $\phi_7$ and $\phi_{10}$ should necessarily originate from physics beyond the SM.
A few observations on the asymmetries help much in simplifying the analysis: ($i$) Due to the dependencies 
of the asymmetries on the Wilson coefficients, it is clear that one can re--phase one of the Wilson coefficients.
For instance, one can choose $\phi_9\equiv 0$ leaving $C_{9}^{eff}(m_{b})$ with its SM phases only. ($ii$) As mentioned
above, the Wilson coefficients $C_7^{eff}(m_b)$ and $C_{10}(m_b)$ cannot develop strong phases from light quark loops
so that $\phi_7$ and $\phi_{10}$ can be chosen to have purely $weak$ character.
 
In the light of analytic derivations as well as particular observations mentioned above, one can investigate 
the dependence of the asymmetries on these $new$ $weak$ phases $\phi_7$ and $\phi_{10}$ to have an estimate 
of their information content. It is conceivable that such an analysis will provide a tool to mark possible 
sources of CP violation beyond the CKM matrix. 

\section{Numerical Estimates}

In this section we present our numerical estimates for the asymmetries
$A_{CP},~P_L,~P_T$ and $P_N$ for $B \rar K^\ast \mu^+ \mu^-$ and $B \rar K^\ast
\tau^+ \tau^-$ decays separately. We take hadronic form factors from Table I and the Wilson coefficients
from Table II. For the remaining parameters we take 
$m_b=4.8~GeV,~m_c=1.35~GeV,~m_B=5.28~GeV,~m_{K^\ast}=0.892~GeV$. 
 
The dilepton invariant mass has the kinematical interval $4 m_{\ell}^{2} \leq q^2 \leq (m_B-m_{K^\ast})^2$
in which the charmonium resonances can be excited. The dominant contribution comes from the three
low--lying resonances $J/\psi, \psi^{'}, \psi^{''}$ in the interval  $8~GeV^2\simlt q^2\simlt
14.5~GeV^2$. In order to minimize the hadronic uncertainties we will discard this subinterval 
in the analysis below by dividing the $q^2$ region to $low$ and $high$ dilepton mass intervals         
\bea
&&\mbox{Region I}:\ \ \ 4 m_{\ell}^{2} \leq q^2 \leq  8~GeV^2,\nnb\\ 
&&\mbox{Region II}:\ \ \ 14.5~GeV^2 \leq q^2 \leq  (m_B-m_{K^\ast})^2,
\eea
where the contribution of the higher resonances do still exists in the second region. 

Due to $1/q^2$ factor in front of $C_{BR}$, in Region I the contribution of the dipole type 
operator dominates. Therefore, asymmetries which involve the differences of the decay rates are
suppressed in Region I compared to ones in Region II. This property will be illustrated in Figs. 1 -- 2
and the remaining analysis for the asymmetries will be performed only for Region II where the asymmetries
are larger.

As mentioned previously, in the model under concern, there are two weak phases: $\phi_7$ and $\phi_{10}$.
However, a close inspection of the CP asymmetry shows that, it is independent of $\phi_{10}$. This
follows from the fact that CP asymmetry can exist only when interference terms involve $strong$ and
$weak$ phases. In this model, similar to SM, there is no interference terms involving $C_{9}^{eff}$ and
$C_{10}$. For this reason $A_{CP}$ is independent of $\phi_{10}$.

First, we illustrate $A_{CP}$ in $\phi_{7}$--$q^{2}$ plane for $B \rar K^\ast \mu^+ \mu^-$
decay for Region I and Region II in Fig. 1 and Fig. 2, respectively. In both figures we take
$\phi_{10}=0$, and as noted above  $A_{CP}$ is already independent of this phase.  In Region I the 
CP asymmetry is practically independent of $q^2$, and becomes maximal for marginal CP violation,
$\phi_{7}=\pi/2$. In Region II, however, the $q^2$ dependence is comparatively enhanced as the 
dominance of dipole coefficient is now reduced. Besides, as figures suggest the CP asymmetry 
in Region II is one order of magnitude larger than in Region I, and  this confirms our expectation above. 

One notes that the average asymmetries could be measured more easily in experiments.
Therefore, from now on we will discuss averaged CP and lepton polarization asymmetries in Region II. The
averaging procedure is defined by 
  
\begin{eqnarray}
\lla Q \rra = \frac{\int_{14.5~GeV^2}^{(m_B-m_{K^\ast})^2} 
Q \frac{d \Gamma}{d q^{2}} d q^{2}}{\int_{14.5~GeV^2}^{(m_B-m_{K^\ast})^2}
\frac{d \Gamma}{d q^{2}} d q^{2}}
\end{eqnarray}
where $Q=P_L, P_N, P_T$ or $A_{CP}$.

Depicted in Fig. 3 (Fig. 4) is the $\phi_7$ dependence of the averaged asymmetries $\lla P_L \rra$, 
$\lla P_T \rra$, $\lla P_N \rra$ and $\lla A_{CP} \rra$ at $\phi_{10} = 0$ ( $\phi_{10} = \pi/2$)
for $B \rar K^\ast \mu^+ \mu^-$ decay. Similarly, Figs. 5 and 6 show the $\phi_7$ dependence of the 
same quantities for $B \rar K^\ast \tau^+ \tau^-$. As noted before, the CP asymmetry depends only 
on $\phi_7$; however, as these figures show clearly among all asymmetries $P_N$ is 
very sensitive to $\phi_{10}$: For $\phi_{10}=0$ ($\phi_{10} = \pi/2$)  $P_N$ is purely positive
(negative). In addition to this, $P_N$ at $\phi_{10} = \pi/2$ is one order of magnitude
larger than that at $\phi_{10} = 0$. This property is valid for both $\mu^+\mu^-$ and  $\tau^+\tau^-$
final states. Besides, since $P_N$ is proportional to the lepton mass, the  $B \rar K^\ast
\tau^+ \tau^-$ decay is much more relevant for its measurement. This sensitivity of $P_N$ on $\phi_{10}$
can be explained as follows: $P_N$ depends on the imaginary part of the bilinear combinations of the 
Wilson coefficients, such as ${\rm Im}\left[C_9^{eff} C_{10}^{\ast}\right]$.  When $\phi_{10}=\pi/2$ (
$\phi_{10}=0$) $C_{10}$ is pure imaginary (real) and therefore ${\rm Im} [C_9^{eff}
C_{10}^{\ast}]=|C_{10}| {\rm Re}[C_9^{eff}]$ (${\rm Im}[C_9^{eff}
C_{10}^{\ast}]=|C_{10}| {\rm Im}[C_9^{eff}]$). Since $|{\rm Re} [C_9^{eff}]| > > |{\rm
Im}[C_9^{eff}]|$, $P_N$ at $\phi_{10}=\pi/2$ is roughly one order of magnitude larger than its
value at $\phi_{10}=0$. Remaining two asymmetries, $P_L$ and $P_T$, are less sensitive to $\phi_{10}$.

In Fig.7 and 8 we present the correlation between $\lla A_{CP}\rra$ and $\lla P_{N}\rra$ for  $B \rar
K^\ast \tau^+ \tau^-$ decay by varying $\phi_7$ from 0 to $2 \pi$ at $\phi_{10}=0$ and
$\phi_{10}=\pi/2$, respectively. For $\mu^+\mu^-$ channel $P_{N}$ is much smaller so we do 
not analyze this case. The SM predictions are given by the intersections of $\lla A_{CP}\rra=0$ line
and the curves themselves. Due to the sign ambiguity of $C_7$ there are two solutions. All other 
points on the curves are generated by the new physics phases. If a simultaneous measurement of  $\lla
A_{CP}\rra$ and $\lla P_{N}\rra$ gives a point on the curve and if this point is distinct from the 
SM prediction then this will be an indication of the new physics contribution. Moreover, such a
simultaneous measurement enables us to determine the sign of the new phases unambiguously. 

\section{Conclusion}
In this work we have adopted a model--independent approach in studying the sensitivity of
the CP and lepton polarization asymmetries to new CP phases. In parcticular, we have taken
the Wilson coefficients being identical to the SM ones except for their phases.
The main result of the present study is that the CP asymmetry and normal lepton polarization
asymmetry are the most sensitive quantities to new sources of weak phases beyond the SM. While $A_{CP}$
is sensitive to $\phi_7$ only, $P_N$ is more sensitive to $\phi_{10}$. Therefore, measurement of these  two
asymmetries can establish the existence or absence of the new sources of CP violation beyond the SM.
Moreover, a simultaneous measurement of the averaged CP and normal polarization asymmetries will
unambiguously determine the sign of the new phases. 

In a specific model such as 2HDM or supersymmetry the Wilson coefficients possess new CP phases 
not found in the SM. The question of how infromative the asymmetries in $B \rar K^\ast \ell^+ \ell^-$
decay about new sources of CP violation in 2HDM model or supersymmetry will be discussed elsewhere.

\newpage
\section*{Figure captions}
{\bf Fig. 1} The dependence of the CP asymmetry $A_{CP}$ 
for  $B \rar K^\ast \mu^-\mu^+$ on $q^2$ and $\phi_7$ 
at $\phi_{10}=0$ for Region I.\\ \\
{\bf Fig. 2} The same as in Fig. 1 but for Region II.\\ \\
{\bf Fig. 3} The dependence of $\lla A_{CP} \rra$,
$\lla P_L \rra$,  $\lla P_T \rra$  and $\lla P_N \rra$ 
for $B \rar K^\ast \mu^-\mu^+$ on $\phi_7$ at $\phi_{10}=0$
for Region II.\\ \\
{\bf Fig. 4} The same as in Fig. 3 but for $\phi_{10}=\pi/2$.\\ \\
{\bf Fig. 5} The same as in Fig. 3 but for  $B \rar K^\ast \tau^-\tau^+$ decay.\\ \\  
{\bf Fig. 6} The same as in Fig. 5 but for $\phi_{10}=\pi/2$.\\ \\
{\bf Fig. 7} The correlation between the averaged CP and normal lepton
polarization asymmetry at $\phi_{10}=0$ for $B \rar K^\ast \tau^-\tau^+$ decay.\\ \\ 
{\bf Fig. 8} The same as Fig. 7 but for $\phi_{10}=\pi/2$.\\ \\

\begin{figure}[H]
\vskip 1.5 cm
    \includegraphics{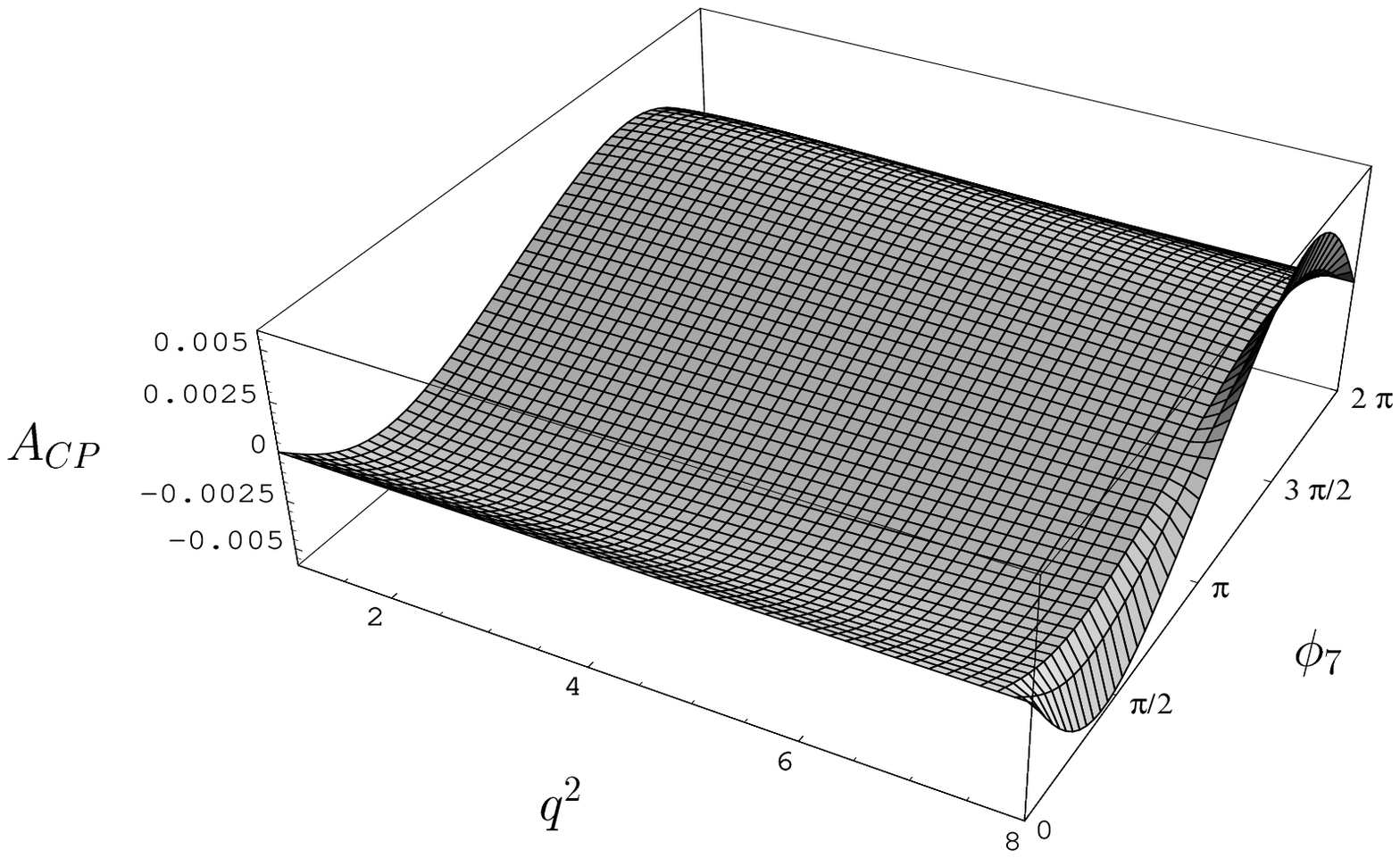}
\vskip 6.7cm   
\caption{}
\end{figure}

\begin{figure}[b]
\vskip 0 cm
    \includegraphics{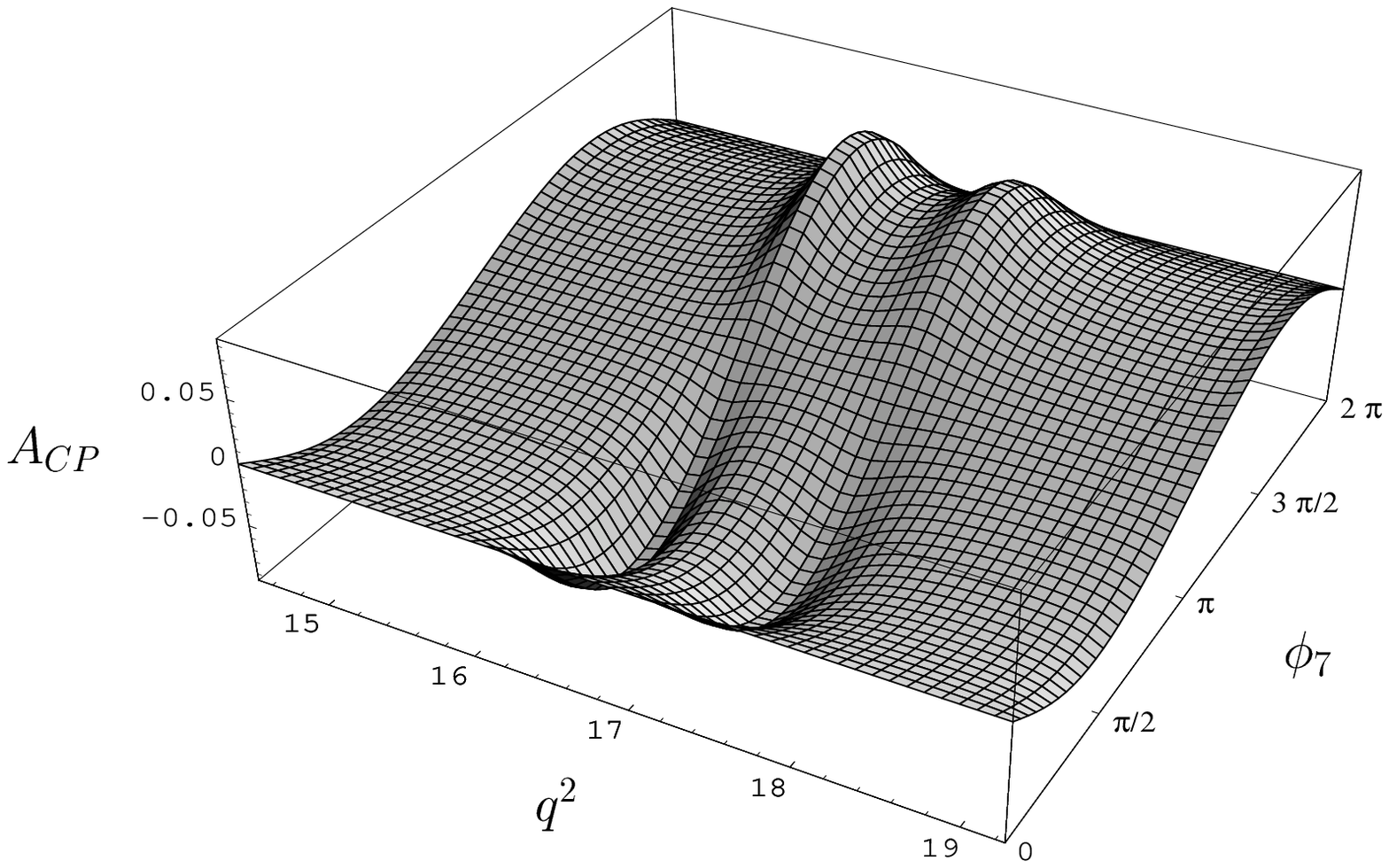}
\vskip 9.5 cm
\caption{ }
\end{figure}

\begin{figure}
\vskip 1.5 cm
    \includegraphics{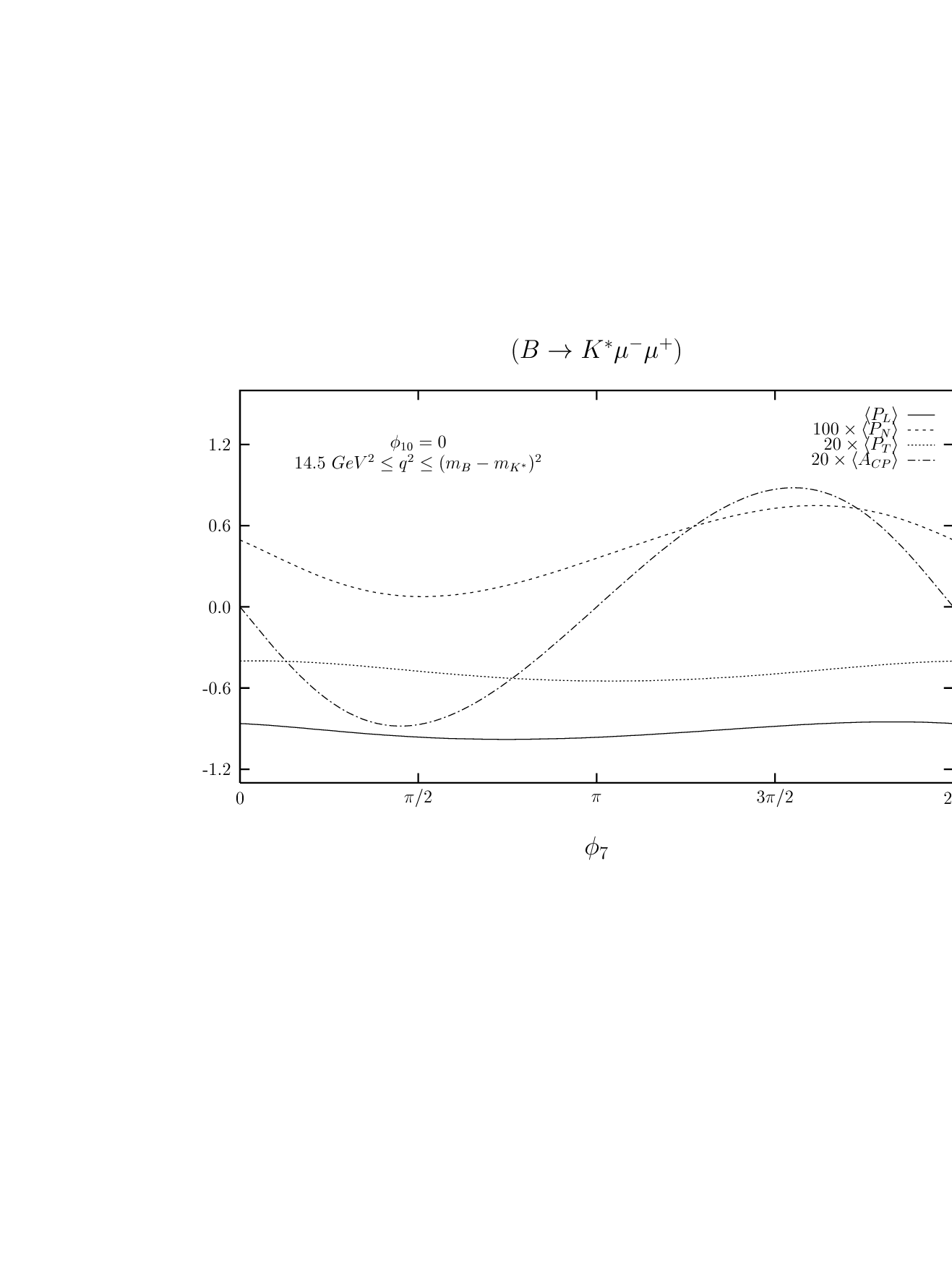}
\vskip 8.4 cm
\caption{}
\end{figure}

\begin{figure}
\vskip 3 cm
    \includegraphics{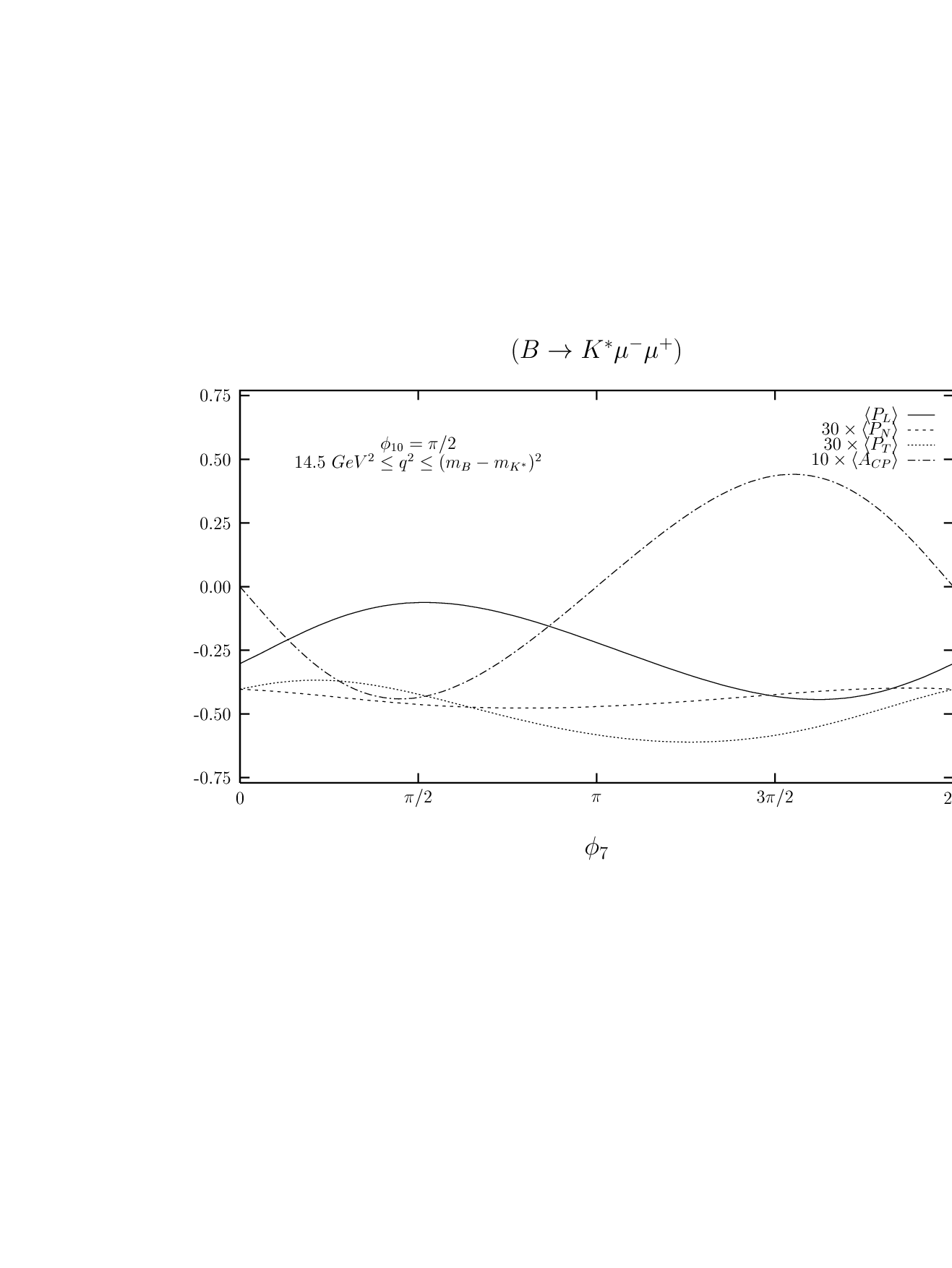}
\vskip 7.4 cm
\caption{}
\end{figure}

\begin{figure}
\vskip 1.5 cm
    \includegraphics{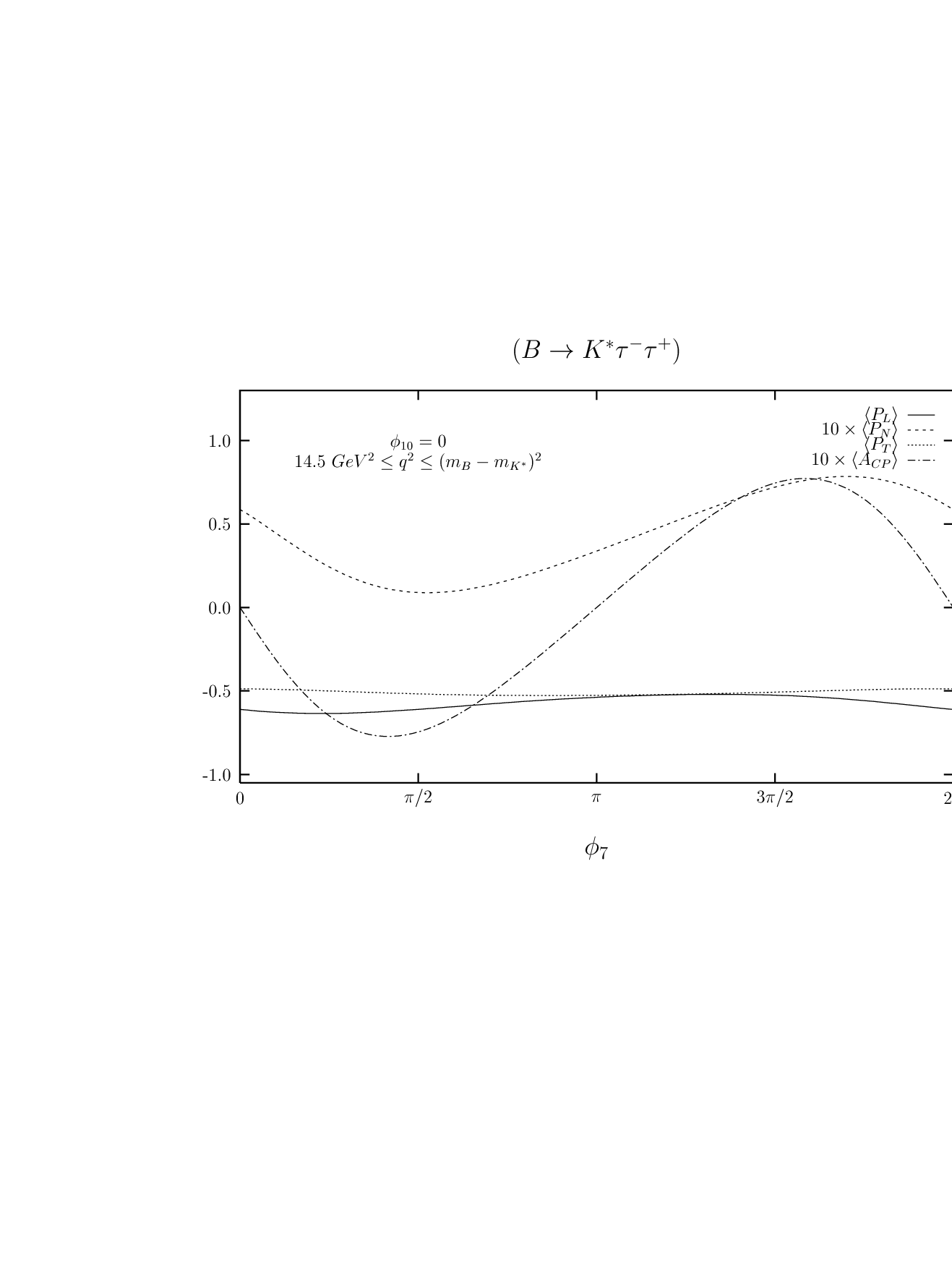}
\vskip 8.4 cm
\caption{}
\end{figure}

\begin{figure}
\vskip 3 cm
    \includegraphics{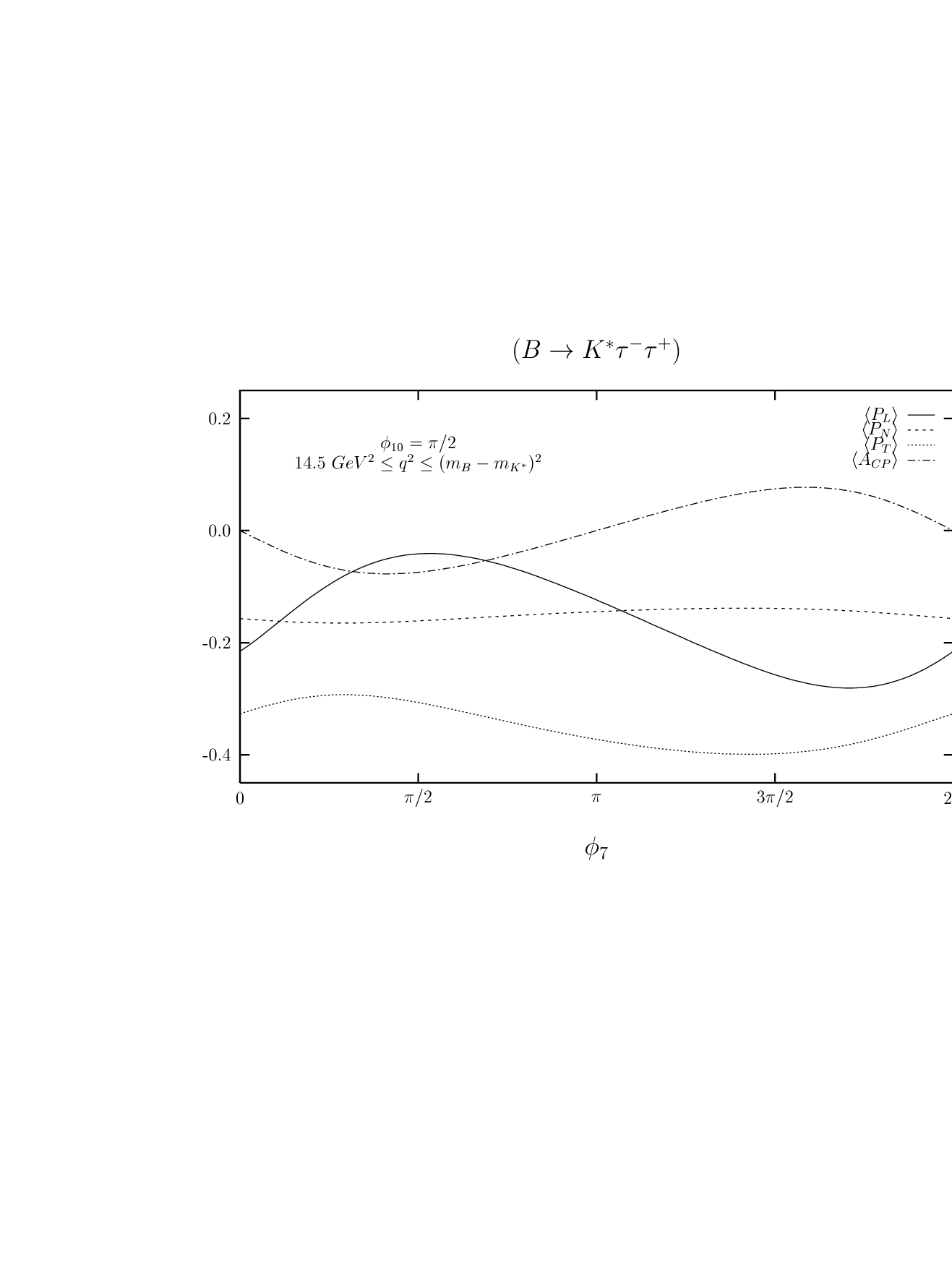}
\vskip 7.4 cm
\caption{}
\end{figure}

\begin{figure}
\vskip 1.5 cm
    \includegraphics{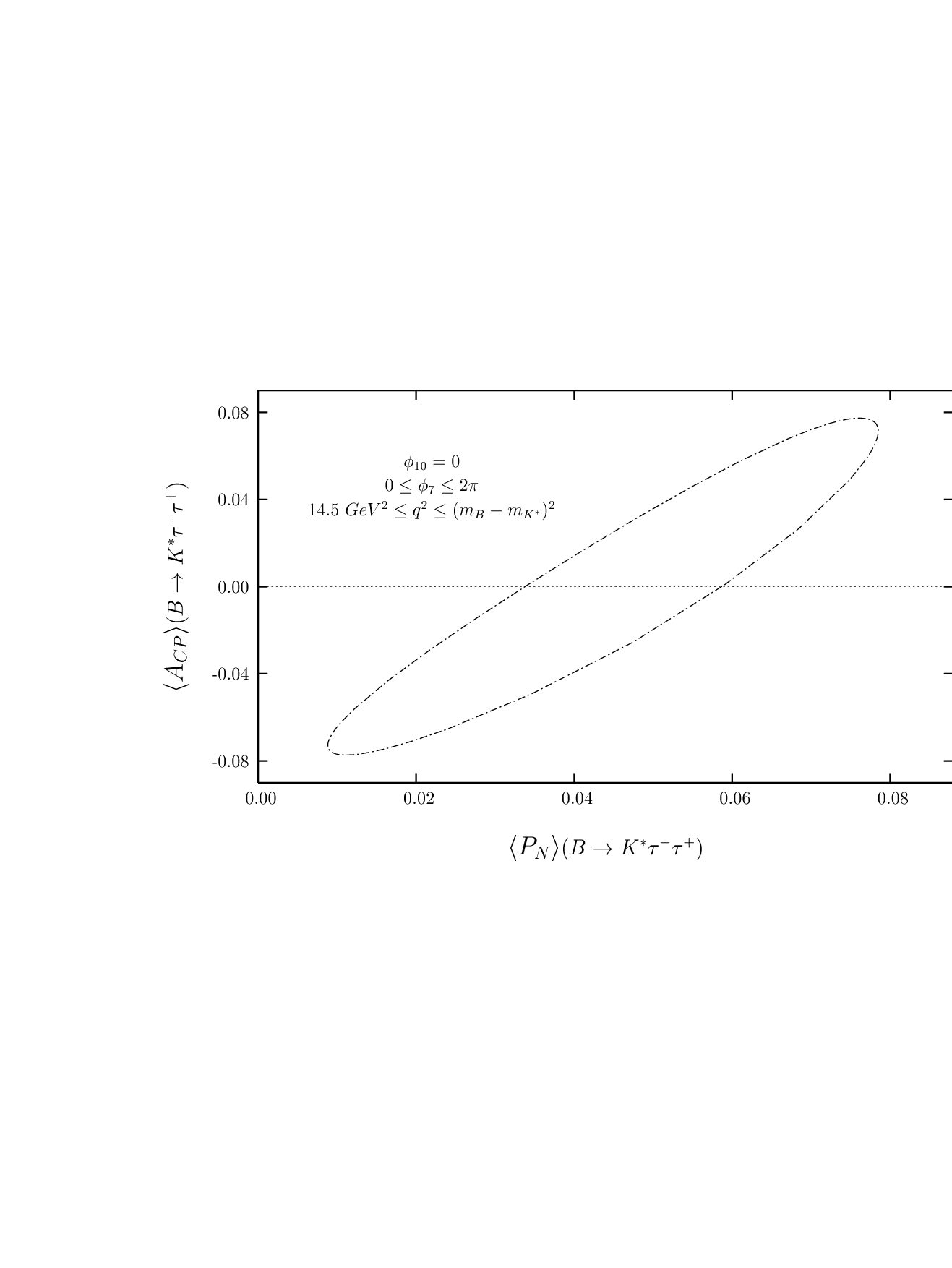}
\vskip 8.4 cm
\caption{}
\end{figure}

\begin{figure}
\vskip 3 cm
    \includegraphics{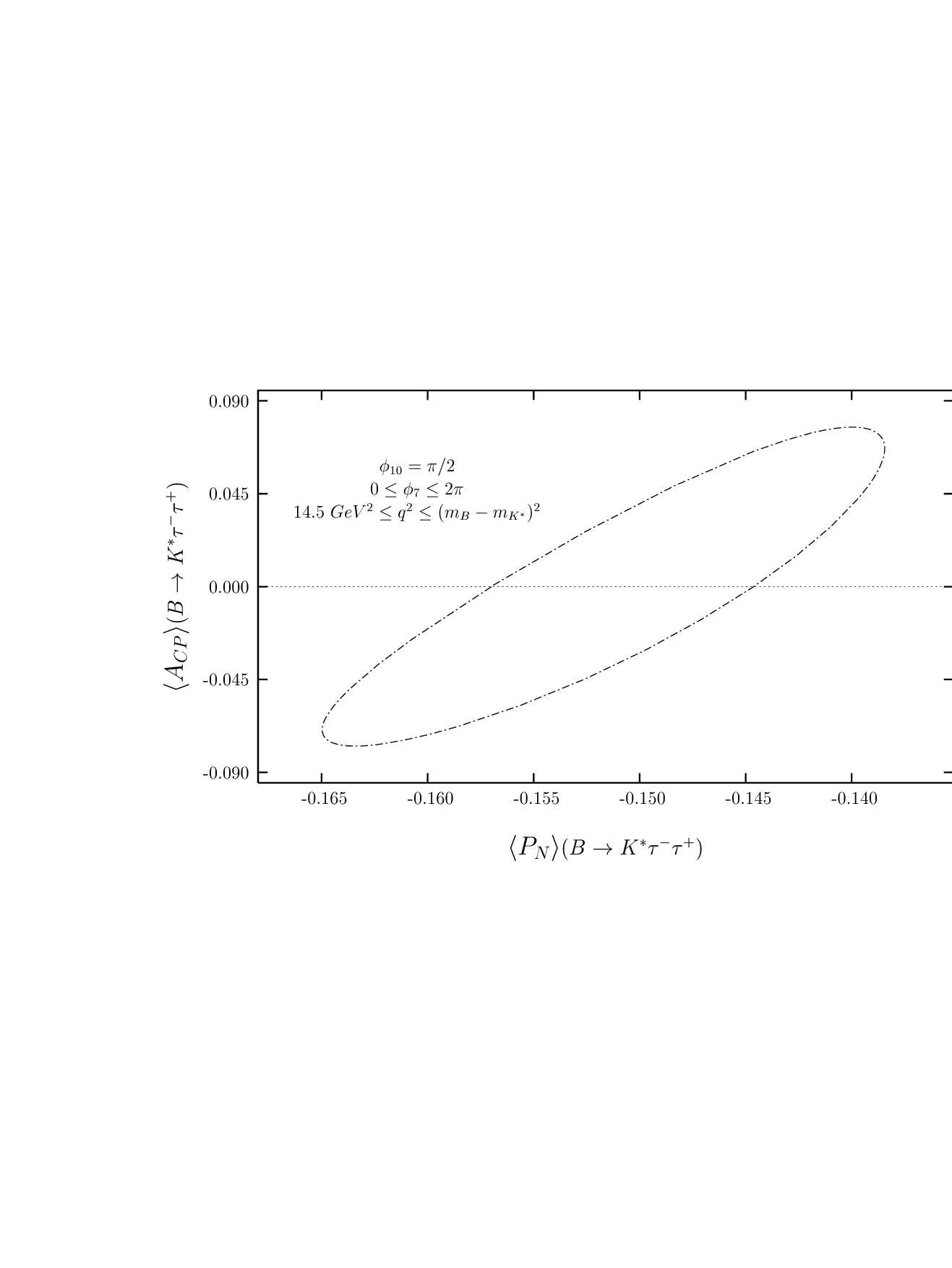}
\vskip 7.4 cm
\caption{}
\end{figure}


\newpage

\begin{thebibliography}{99}
\bibitem{kaon} J.H. Christenson, J.W. Cronin, V.L. Fitch, R. Turlay,
Phys. Rev. Lett. {\bf 13} (1964) 138; KTeV Collaboration,
Phys. Rev. Lett. {\bf 83} (1999) 22.

\bibitem{CLEO} M. S. Alam {\it at. al}, CLEO Collaboration,
Phys. Rev. Lett. {\bf 74} (1995) 2885;\\
R. Ammar {\it at. al}, CLEO Collaboration,
Phys. Rev. Lett. {\bf 71} (1993) 674.

\bibitem{R2} C. Q. Geng and P. C. Kao,
Phys. Rev. {\bf D54} (1996) 5636.

\bibitem{R3} T. M. Aliev, H. Koru, A. \"{O}zpineci and M. Savc{\i},
Phys. Lett. {\bf 400B} (1997) 194; 
T. M. Aliev, M. Savc{\i}, A. \"{O}zpineci and H. Koru,
J. Phys. G {\bf 24} (1998) 49; T. M. Aliev, D. A. Demir, E. Iltan, N. K. Pak,
Phys. Rev. {\bf D54} (1996) 851.

\bibitem{R4} F. Kr\"{u}ger and L. M. Sehgal
 Phys. Rev. {\bf D55} (1997) 2799.

\bibitem{R5} S. Fukae, C. S. Kim, T. Yoshikawa,
hep-ph/9908229 (1999).

\bibitem{R6} T. M. Aliev, C. S. Kim, Y. G. Kim,
hep-ph/9910511 (1999). 

\bibitem{R7} A. Ali, P. Ball, L. T. Handako and G. Hiller,
hep-ph/9910221 (1999). 


\bibitem{R9} S. Fukae, C. S. Kim, T. Morozumi and T. Yoshikawa,
Phys. Rev.  {\bf D59} (1999) 074013.

\bibitem{R10} P. Colangelo, F. De  Fazio, P. Santorelli and E. Scrimieri,\\
Phys. Rev. {\bf D53} (1996) 3672; (E) {\bf D57} (1998) 3186.

\bibitem{R11} R. Casalbuoni, A. Deandrea, N. Di Bartolomeo, R Gatto, G. Nardulli, \\
Phys. Lett. {\bf B312} (1993) 315.

\bibitem{R12} W. Jaus and D. Wyler,
Phys. Rev. {\bf D41} (1990) 3405.

\bibitem{R13} W. Roberts,
Phys. Rev. {\bf D54} (1996) 863.

\bibitem{R14} T. M. Aliev, A. \"{O}zpineci and M. Savc{\i},
Phys. Rev. {\bf D55} (1997) 7059.

\bibitem{R15}  P. Ball and V. M. Braun,
Phys. Rev. {\bf D58} (1998) 094016.

\bibitem{R16}  P. Ball, JHEP (1998) 005. 

\bibitem{erhanabi} T.M. Aliev, E.O. Iltan, Phys. Lett. {\bf 451B} (1999) 175.

\bibitem{cdf1} T. Affolder {\it et. al.}, CDF Collaboration, Phys. Rev. Lett. {\bf 83} (1999) 3378. 

\bibitem{LEP} G. Abbiendi {\it et. al.}, OPAL Collaboration, hep-ex/9908002.   

\bibitem{R17}  M. Misiak, 
Nucl. Phys. {\bf B393} (1993) 23; (E) {\bf B439} (1995) 461.

\bibitem{R18} A. J. Buras and M. M\"{u}nz,
Phys. Rev. {\bf D52} (1995) 186.

\bibitem{bw} N. G. Deshpande, J. Trampetic and K. Panose, Phys. Rev. {\bf D39} (1989) 1461;
C. S. Lim, T. Morozumi and A. Sanda, Phys. Lett. {\bf 218B} (1989) 343.

\end{thebibliography}
\end{document}